\documentclass[letterpaper,titlepage,11pt]{article}

\pdfoutput=1

\usepackage{amssymb,amsmath,amsfonts}
\usepackage{bm}
\usepackage{epsfig}
\usepackage{ccaption}
\usepackage{graphicx}
\usepackage{float}
\usepackage{breqn}

\usepackage[
      colorlinks=true,
      linkcolor=blue,
      urlcolor=blue,
      filecolor=black,
      citecolor=red,
      pdfstartview=FitV,
      pdftitle={},
        pdfauthor={Roberto Emparan, Raimon Luna, Marina Martinez, Ryotaku Suzuki, Kentaro Tanabe},
        pdfsubject={},
        pdfkeywords={},
        pdfpagemode=None,
        bookmarksopen=true
      ]{hyperref}

\def\clock{{\count0=\time
           \divide\count0 60
           \ifnum\count0<10 0\fi\the\count0
           \multiply\count0 -60 \advance\count0 \time
           :\ifnum\count0<10 0\fi \the\count0
         }}
\newcommand{\timestamp}{{\small\vbox{\hbox{\tt\jobname.tex}
\hbox{\the\day/\the\month/\the\year, \clock}}}}

\newcommand{\nn}{\nonumber}
\newcommand{\ie}{{\it i.e.,\,}}
\newcommand{\eg}{{\it e.g.,\,}}

\newcommand{\lp}{\left(}
\newcommand{\rp}{\right)}

\newcommand{\beq}{\begin{equation}}
\newcommand{\eeq}{\end{equation}}
\newcommand{\bea}{\begin{eqnarray}}
\newcommand{\eea}{\end{eqnarray}}
\newcommand{\beqa}{\begin{eqnarray}}
\newcommand{\eeqa}{\end{eqnarray}}

\newcommand{\sR}{\mathsf{R}}

\newcommand{\ord}[1]{{\mathcal O}\lp #1\rp}

\numberwithin{equation}{section}

\begin{document}

\begin{titlepage}
\rightline{AP-GR-143, OCU-PHYS-475}
\vskip 2cm
\centerline{\LARGE \bf Phases and Stability of }
\medskip
\centerline{\LARGE \bf Non-Uniform Black Strings}
\vskip 1.2cm
\centerline{\bf Roberto Emparan$^{a,b}$, Raimon Luna$^{b}$, Marina Mart{\'\i}nez$^{b,c}$,} 
\centerline{\bf Ryotaku Suzuki$^{d}$, Kentaro Tanabe$^{e}$\footnote{Until March 2017}}
\vskip 0.5cm
\centerline{\sl $^{a}$Instituci\'o Catalana de Recerca i Estudis
Avan\c cats (ICREA)}
\centerline{\sl Passeig Llu\'{\i}s Companys 23, E-08010 Barcelona, Spain}
\smallskip
\centerline{\sl $^{b}$Departament de F{\'\i}sica Qu\`antica i Astrof\'{\i}sica, Institut de
Ci\`encies del Cosmos,}
\centerline{\sl  Universitat de
Barcelona, Mart\'{\i} i Franqu\`es 1, E-08028 Barcelona, Spain}
\smallskip
\centerline{\sl $^{c}$Institute for Theoretical Physics, KU Leuven, Celestijnenlaan 200D, 3001 Leuven, Belgium}
\smallskip
\centerline{\sl $^{d}$Department of Physics, Osaka City University, Osaka 558-8585, Japan}
\smallskip
\centerline{\sl $^{e}$Department of Physics, Rikkyo University,
Toshima, Tokyo 171-8501, Japan}
\smallskip

\vskip 1.2cm
\centerline{\bf Abstract} \vskip 0.2cm 
\noindent 
We construct solutions of non-uniform black strings in dimensions from $D\approx 9$ 
all the way up to $D=\infty$, and investigate their thermodynamics and dynamical stability. Our approach employs the large-$D$ perturbative expansion beyond the leading order, including corrections up to $1/D^4$. Combining both analytical techniques and relatively simple numerical solution of ODEs, we map out the ranges of parameters in which non-uniform black strings exist in each dimension and compute their thermodynamics and quasinormal modes with accuracy. We establish with very good precision the existence of Sorkin's critical dimension and we prove that not only the thermodynamic stability, but also the dynamic stability of the solutions changes at it. 

\end{titlepage}
\pagestyle{empty}
\small
\normalsize
\newpage
\pagestyle{plain}
\setcounter{page}{1}

\section{Introduction and Summary}

The instability of black strings and black branes discovered in \cite{Gregory:1993vy,Gregory:1994bj} is an instance of the spontaneous breakdown of a translational symmetry, a phenomenon that is widely present in many other areas of physics (for instance, in hydrodynamics and in materials science). It is of interest in itself for a deeper understanding of the dynamics of spacetime, but in addition, through holographic dualities, it can be connected to the spontaneous development of inhomogeneity in non-gravitational systems.

Establishing the existence of an instability of the translationally invariant, uniform black string (UBS) solution is relatively easy, as it only involves the study of linearized mode perturbations: a mode with negative imaginary frequency signals an instability. At the threshold of the instability, when the purely imaginary frequency changes sign, the zero-frequency perturbation results in a time-independent, sinusoidal spatial modulation along the black string horizon. This indicates the beginning of a branch of static non-uniform black strings (NUBS) \cite{Gubser:2001ac,Wiseman:2002zc}. It is much harder to identify the stability properties of the solutions in this branch, to find what is the most probable endpoint of the initial instability, and, eventually, to determine all the possible phases of the system.

In general, all these issues about the existence and the local and global stability of solutions depend on the values of the parameters of the system, such as its charges and the number of dimensions. Typically the linear instability is present only within certain parameter ranges, but also the non-linear evolution, and its most probable endpoint, depend in general on what region of parameter space the initial system lies in. For black strings and black branes these details are not fully understood; the non-linear regime requires the numerical solution of Einstein's equations, both to find highly non-uniform static phases, and to determine their stability and time-evolution (see \cite{Kol:2004ww,Harmark:2007md,Horowitz:2012nnc} for early reviews and \cite{Kalisch:2018efd} for a very recent one). 
Studies so far have mostly focused on the properties of neutral, asymptotically flat (in transverse directions) black strings in different numbers of dimensions $D$.\footnote{See \cite{Rozali:2016yhw,Dias:2017coo} for recent work on non-uniform black membranes.}   The latter is possibly the simplest and most natural parameter that can be varied to explore this problem. Indeed, since a number $D-3$ of the dimensions are fixed in the shape of a $S^{D-3}$, the equations reduce to a three-dimensional dynamics where $D$ is a parameter that can be varied continuously. Hence, in this article we focus on neutral black strings in different $D$, naturally allowing non-integer values of it.

The initial work of Gregory and Laflamme (GL) \cite{Gregory:1993vy,Gregory:1994bj} established that uniform black strings are unstable up to $D=10$ and very likely in all $D\geq 5$. Then in ref.~\cite{Sorkin:2004qq} Sorkin argued that the nature of the symmetry-breaking phase transition changes, from first to second order, at a critical dimension $D_* \simeq 13.5$.\footnote{See \cite{Park:2004zr} for a heuristic argument for the presence of this critical dimension.} Numerical constructions of NUBS in the non-linear regime were first performed in \cite{Wiseman:2002zc} and then followed up in \cite{Kleihaus:2006ee,Sorkin:2006wp,Figueras:2012xj,Kalisch:2015via,Kalisch:2016fkm,Kalisch:2017bin}. In particular ref.~\cite{Figueras:2012xj} scanned the thermodynamic properties of static black strings up to $D=15$, and revealed non-trivial detailed features in the phase space around the critical dimension. All of these studies of static configurations involved numerical solution of the equations. Results on the dynamical evolution of black strings are much sparser, see \cite{Lehner:2010pn} in $D=5$, and \cite{Emparan:2015gva,Rozali:2016yhw,Miyamoto:2017ozn} in $D\to\infty$.

\subsubsection*{Approach and summary of results}

Since $D$ can be treated in this system as a continuous variable, one can perform a perturbative analysis around the limit $D\to\infty$. In this article we use the large-$D$ expansion of black brane dynamics \cite{Asnin:2007rw,Emparan:2013moa} to investigate all these problems through a combination of analytical and (relatively simple) numerical methods.\footnote{In addition to those mentioned above, references relevant to the large-$D$ approach to these issues are \cite{Kol:2004pn,Emparan:2014aba,Emparan:2015rva,Emparan:2015hwa,Suzuki:2015axa,Emparan:2016sjk,Bhattacharyya:2015fdk,Dandekar:2016jrp,Dandekar:2016fvw,Herzog:2017qwp}.} 

The large-$D$ approach to the construction of static NUBS was started in \cite{Emparan:2015hwa} and then significantly extended in \cite{Suzuki:2015axa}. In particular, in the latter article an analytical calculation showed the presence of Sorkin's critical dimension, pinning it down to a value $D_*\simeq 13.5$ uncannily close to the numerical result of \cite{Sorkin:2004qq}. Ref.~\cite{Emparan:2015gva} obtained the leading order (LO) large-$D$ effective dynamical equations for the system, and then performed time evolutions which showed that when $D\to\infty$ the instability ends at inhomogeneous stable black strings.  Ref.~\cite{Rozali:2016yhw} obtained dynamical equations to next-to-leading order (NLO) (independently of our ongoing work at the time) and used them to scan the phase space and stability of solutions.

Here we extend and greatly refine these studies to cover most of the main properties of NUBS over virtually the entire range of possible dimensions, above and below the critical dimension. We expand the equations in $1/D$ successively including LO, NLO, and even up to 4NLO corrections (\ie $1/D^4$).  With this, we manage to:
\begin{itemize}
\item Construct inhomogeneous black strings, analytically at weak non-uniformity, and numerically at larger non-uniformity, and compute their thermodynamical properties.
\item Show analytically that weakly non-uniform black strings are dynamically stable above, and unstable below, a critical dimension $D_*= 13.6$, coincident with the thermodynamical critical dimension.
\item Numerically evolve in time unstable black strings, until they either stabilize on NUBS, or else---if they are thin enough, or if $D$ is low enough---the evolution eventually breaks down.
\end{itemize}

With these results we produce phase diagrams of mass, entropy and relative binding energy (or tension per unit mass and unit length \cite{Harmark:2003dg}) as functions of the temperature. Some of these diagrams can be compared to those that ref.~\cite{Figueras:2012xj} obtained at several specific values of $D$  (with numerical methods rather more sophisticated than ours). The qualitative agreement is remarkably good: we reproduce all the main features that \cite{Figueras:2012xj} identified for how NUBS phases branch off from the uniform phases. In particular, we can establish the presence of thermodynamically and dynamically stable NUBS below the critical dimension if they are inhomogeneous enough. All these features can be seen in our analytically-constructed weakly non-uniform black strings. Moreover, the quantitative agreement with \cite{Figueras:2012xj} is excellent above and even below the critical dimension.

\subsubsection*{Range of existence of NUBS}

We also perform a study of highly non-uniform black strings in increasing number of dimensions, in order to establish when they can be expected to transition into localized black holes. We do this by a combination of methods---dynamical evolution leading to large non-uniformities until the numerical evolution breaks down, and studies of static solutions that develop identifiable pathologies (specifically, negative tensions) when the non-uniformity grows too large.\footnote{Ref.~\cite{Rozali:2016yhw} made similar studies to NLO.} In dynamical evolutions we aim mostly at qualitative information, for which NLO effects seem enough\footnote{At times we have gone to NNLO, to confirm our conclusions.}, but in the construction of static solutions we strive for accuracy comparable to previous studies and thus we include up to 4NLO corrections.


\begin{figure}[t]
\begin{center}
\includegraphics[width=.495\textwidth]{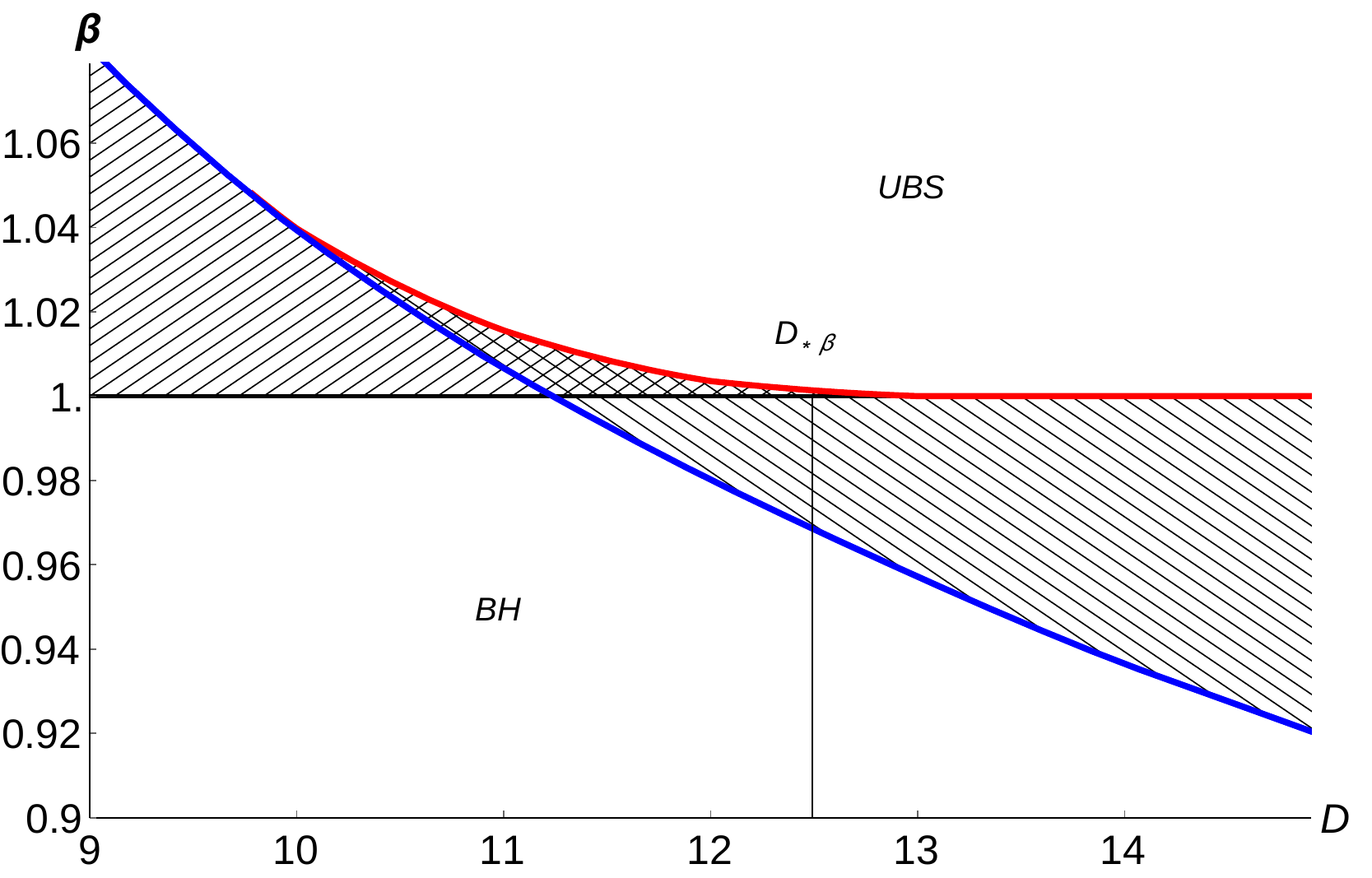}
\includegraphics[width=.495\textwidth]{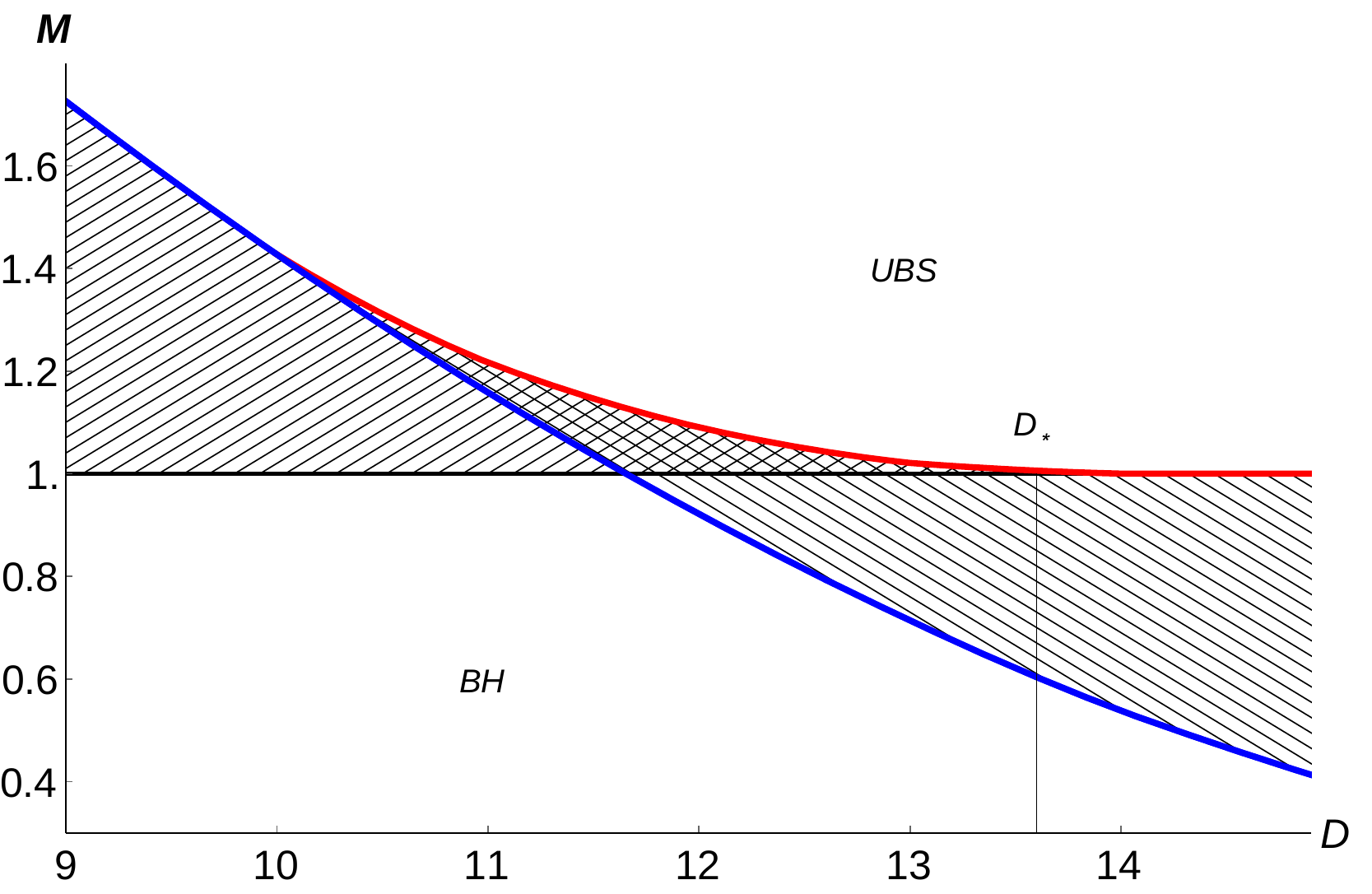}
\caption{\small Range of existence (hatched) of non-uniform black strings in a compact circle of fixed length, as a function of the spacetime dimension $D$ and the inverse temperature $\bm{\beta}$ or the mass $\mathbf{M}$ (results obtained numerically at 4NLO). The vertical axes are normalized relative to the value for a uniform black string at the threshold of the GL instability, so the lines $\bm{\beta},\mathbf{M}=1$ separate UBS into stable above the line, and unstable below the line. $D_*\simeq 13.6$ is Sorkin's critical dimension \cite{Sorkin:2004qq}, which separates the dimensions in which weak non-uniformity makes $\mathbf{M}$ increase $(D<D_{*})$ or decrease $(D>D_{*})$. $D_{*\beta}\simeq 12.5$ plays a similar role for $\bm{\beta}$. The blue curve marks NUBS that reach zero tension, as a proxy to the transition to localized black holes (BH). The red curve indicates local maxima of $\bm{\beta}$ or $\mathbf{M}$, reached as the non-uniformity increases when $9\lesssim D<D_{*\beta},D_{*}$. The doubly-hatched region is a fold in phase space, where two NUBS phases coexist with the same $\bm{\beta}$ or the same $\mathbf{M}$, but differ in their entropies. For $D\lesssim 10$, NUBS branches terminate before any local maximum of $\bm{\beta}$ or $\mathbf{M}$ is reached. BH phases (not studied in this paper) coexist and can dominate thermodynamically over NUBS in regions of $\bm{\beta},\mathbf{M}>1$ at low $D$. 
The blue curves behave asymptotically at large $D$ like $\bm{\beta}\sim D^{-1/4}$ and $\mathbf{M}\sim D^{-D/4}$.
}\label{fig:phases}
\end{center}
\end{figure}

Collecting all this information we produce a diagram, see fig.~\ref{fig:phases}, for the expected range of existence of NUBS in a circle of fixed length, in the inverse temperature $\bm{\beta}$ (canonical ensemble) and in the mass $\mathbf{M}$ (microcanonical ensemble), as a function of the number of dimensions $D$. We normalize $\bm{\beta}$ and $\mathbf{M}$ relative to their values for a UBS at the threshold of the GL instability in the same circle. To get oriented, as the value of $\bm{\beta}$, or $\mathbf{M}$, decreases the UBS become thinner. UBS exist for all values of $\bm{\beta}$ and $\mathbf{M}$ in all $D\geq 5$, and are locally (GL) stable for $\bm{\beta},\,\mathbf{M}>1$ and unstable for $\bm{\beta},\,\mathbf{M}<1$.

In the diagram we include all NUBS phases, whether they are stable or not, either linearly, non-linearly, or thermodynamically. The blue line is a proxy for the transition of NUBS to localized black holes. This merger point is outside the applicability of the large-$D$ approach to NUBS, so instead we estimate it as the point at which NUBS
become so inhomogeneous that they develop negative tension. Therefore, the blue line should only be taken as a semiquantitative boundary. Up to this caveat, we expect fig.~\ref{fig:phases} to capture qualitatively and quantitatively the properties of NUBS in dimensions around the critical value and higher. In particular, we expect that the asymptotic scaling that we find for the lower boundaries, 
\beq\label{bluelines}
\bm{\beta}_{\textrm{min}}\sim D^{-1/4}\,,\qquad \mathbf{M}_{\textrm{min}}\sim D^{-D/4}\,,
\eeq
is robust for very large $D$. In lower dimensions the quantitative accuracy worsens---hence we only extend the diagram down to $D\approx 9$---, but nevertheless the qualitative features do not seem to differ.

Besides UBS and NUBS it is also possible to have black holes (BH) localized in the compact circle. Less is known in the literature about these phases, except when they are either very small and far from merging into a NUBS \cite{Harmark:2003yz,Gorbonos:2004uc}, or else at relatively low values of $D$ \cite{Kudoh:2004hs,Headrick:2009pv,Kalisch:2017bin}. They are expected to be the dominant phase in the canonical ensemble in the regions below the curves of existence of NUBS. However, there are also regions of the plane where they coexist with NUBS, and in which they may dominate thermodynamically over the latter. In particular, this is expected to be the case in low dimensions up to values of $\bm{\beta}$ and $\mathbf{M}$ larger than 1. However, the details of what are the dominant phases, either at fixed temperature or at fixed mass, can be fairly complicated and require information about localized black holes that we have not explored in this article.

\subsubsection*{Outline}

Section~\ref{sec:BSdyn} introduces the basic framework: the large-$D$ effective theory, the appearance of NUBS, and the calculation of physical magnitudes. Section~\ref{sec:pertstatic} describes the analytic construction of NUBS in an expansion for small non-uniformity. This allows the identification of the thermodynamic critical dimensions $D_*$ and $D_{*\beta}$. Section~\ref{sec:dyn} studies the dynamical stability of perturbative NUBS, in particular their quasinormal spectrum, and establishes, for the first time and analytically, the existence of quasinormal modes that change stability at the critical dimension $D_*$. We also discuss a change of local dynamical and thermodynamical stability at fixed $D<D_*$ when the non-uniformity is large enough. Section~\ref{sec:nums} describes the numerical investigation of NUBS in different dimensions: their range of existence and physical properties, and the time evolution of initially unstable UBS. We conclude in section~\ref{sec:fin} with a discussion of open ends for future investigation.

We have striven to present our main ideas and results without drowning the reader in the explicit results to high orders in the $1/D$ expansion, as well as high orders in an expansion on the non-uniformity of the black strings. These equations can be very lengthy, so for the most part we put them in appendices, or, when they are even too long for that, in ancillary \textsl{Mathematica} files.

\section{Black string dynamics at large $D$}\label{sec:BSdyn}

We begin by briefly recalling the derivation of the effective equations for the dynamics of large-$D$ black strings, and how they easily reveal the appearance of the non-uniform solutions \cite{Emparan:2015gva,Emparan:2016sjk,Dandekar:2016jrp}. The ansatz for the metric is
\begin{equation}\label{metansatz}
ds^2 = -A dt^2 -2 (u_t dt + u_z dz)dr -2 C dz dt + G dz^2 + r^2 d\Omega_{n+1},
\end{equation}
where
\beq
n=D-4
\eeq
will be our large expansion parameter.
Then we expand all functions in $1/n$ as
\begin{equation}\label{funcexpand}
A=\sum_{k \geq 0} \frac{A^{(k)}(t,z,r)}{n^k},\qquad C=\sum_{k \geq 0} \frac{C^{(k)}(t,z,r)}{n^{k+1}}, \qquad G=\frac{1}{n}\left(1+\sum_{k \geq 0} \frac{G^{(k)}(t,z,r)}{n^{k+1}}\right),
\end{equation}
\begin{equation}
u_t=\sum_{k \geq 0} \frac{u_t^{(k)}(t,z,r)}{n^k}, \qquad u_z=\sum_{k \geq 0} \frac{u_z^{(k)}(t,z,r)}{n^{k+1}}.
\end{equation}
We also introduce the radial variable $\sR=r^n$, which is adequate for the region near the horizon of the black string. Then the solution to Einstein's equations at leading order in $1/n$ is easily found to be 
\begin{equation}\label{LOsoln}
A^{(0)} = 1-\frac{m_0(t,z)}{\sR},\qquad  C^{(0)} = \frac{p_0(t,z)}{\sR},\qquad G^{(0)} = \frac{p_0^2(t,z)}{m_0(t,z)\sR},
\end{equation}
\begin{equation}
u_t^{(0)} = -1,\qquad  u_z^{(0)} = \text{const}.
\end{equation}
The functions $m_0(t,z)$ and $p_0(t,z)$, which can be regarded as collective variables for the energy and momentum density along the string\footnote{The momentum density is actually $p_0-\partial_z m_0$ \cite{Emparan:2016sjk}.}, must obey $\sR$-independent constraints of the form
\begin{equation}\begin{split}\label{LOeqs}
&\partial_t m_0-\partial_z^2 m_0 = - \partial_z p_0,\\
&\partial_t p_0-\partial_z^2 p_0 = \partial_z \left(m_0-\frac{p_0^2}{m_0}\right).
\end{split}\end{equation}
Any solution to these equations gives a (generically time-dependent) black string solution of Einstein's theory to leading order in $1/n$. Therefore, to this order, these are the effective equations for the dynamics of the black string. One can also obtain corrections to these equations to higher orders in $1/n$. They are rather lengthy, so we give them in ancillary \textsl{Mathematica} files.

For now it is convenient to choose our length units so that the horizon is at $r=1$ in the uniform black string solution---later we will set units differently. Now take this solution, $m_0=1$, $p_0=0$, and perturb it slighty in the form
\beq\label{linpert}
m_0=1+\delta m \, e^{\Omega t+ikz}\,,\qquad p_0=\delta p  \, e^{\Omega t+ikz}\,.
\eeq
Observe that since the proper length along the black string in \eqref{funcexpand} is $\sqrt{G}\,z=z/\sqrt{n}+\ord{n^{-3/2}}$, then the physical wavenumber is $\sqrt{n}k$.

Linearizing the equations \eqref{LOeqs} in $\delta m$ and $\delta p$ we obtain the spectrum
\beq\label{LOspectrum}
\Omega=\pm k (1\mp k)\,.
\eeq
We see that whenever $0\leq |k|<1$, the perturbation grows, $\Omega>0$, so the black string is unstable. This is the Gregory-Laflamme instability in the large-$D$ limit. 
The wavenumber $k_{GL}=1$ corresponds to the threshold of the instability, where $\Omega=0$. This critical wavenumber can be calculated to higher orders in $1/n$ using the equations in the ancillary \textsl{Mathematica} files. To NNLO we find
\begin{equation}\label{kGLNNLO}
k_{GL} = 1-\frac{1}{2n}+\frac{7}{8n^2}
+\mathcal{O}\left(\frac{1}{n^3}\right)\,.
\end{equation} 
This result was obtained earlier using a linear perturbation analysis in \cite{Asnin:2007rw}, and then extended  with similar methods to 4NLO in \cite{Emparan:2015rva}. We have reobtained the latter result, see \eqref{kGL4NLO}, by linearizing the 4NLO effective equations.

For $k=k_{GL}$ the uniform string admits a small static ($\Omega=0$) sinusoidal perturbation\footnote{Without loss of generality we consider henceforth $k\geq 0$.}. This signals the appearance of a branch of non-uniform black strings. The rest of this article is devoted to understanding the properties of these solutions: how they extend beyond the linear approximation; what their mass, entropy, and other magnitudes are; and whether they are stable or not, both dynamically and thermodynamically, in different number of dimensions. We will also follow the time evolution of the instability into the deeply non-linear regime. In our study we will combine analytic and numerical techniques.

\subsection{Physical magnitudes}

Given a black string solution of the form \eqref{metansatz}, we can compute its physical properties in a conventional manner. For the sake of clarity we give explicit expressions only to LO. Again, higher order results are too lengthy even to show in an appendix, so we collect them in the ancillary files. 

We expand the solution in the asymptotic region for $1\ll \sR\ll e^n$ \footnote{The upper bound guarantees that the calculations remain within the large-$n$ near-horizon zone \cite{Emparan:2013moa}. The matching to the far-zone is explained in \cite{Suzuki:2015axa}.}. We define functions $m(t,z)$ and $p(t,z)$ as the monopolar terms in $A$ and $C$, \ie\
\beqa\label{asympexp}
A&=&1-\frac{m(t,z)+\ord{\ln\sR/n}}{\sR}+\ord{\sR^{-2}}\,,\nn\\
C&=&\frac{p(t,z)+\ord{\ln\sR/n}}{\sR}+\ord{\sR^{-2}}\,.
\eeqa
These are expanded in powers of $1/n$, 
\beq\label{nested}
m(t,z)=\sum_{j=0}^\infty \frac{m_j(t,z)}{n^j}\,,\qquad p(t,z)=\sum_{j=0}^\infty \frac{p_j(t,z)}{n^j}\,.
\eeq

It is now straightforward to compute the quasilocal stress-energy tensor in the  asymptotic region,
\beqa
T_{tt}&=&\frac{\Omega_{n+1}}{16\pi G}(n+1)\lp m(t,z)+\ord{\frac1n}\rp\,,\\
T_{tz}&=&-\frac{\Omega_{n+1}}{16\pi G}\lp p(t,z)-\partial_z m(t,z)+\ord{\frac1n}\rp\,,\\
T_{zz}&=&-\frac{\Omega_{n+1}}{16\pi G}\frac1n \lp m(t,z)-\frac{p^2(t,z)}{m(t,z)}+\partial_z p(t,z)-\partial_t m(t,z)+\ord{\frac1n}\rp\,.
\eeqa
Note that the terms $\propto \ln\sR/\sR$ in the expansion \eqref{asympexp} only contribute to the stress-energy tensor at order $1/n$ and higher. The results up to order $1/n$ for the stress tensor of static solutions and for all other quantities in this section can be found in appendix~\ref{app:thermo}. 

For time-independent solutions we find convenient to define the  mass density along the string (adequately rescaled to absorb $n$-dependent prefactors\footnote{Note that in $M$ we factor out $(n+1)$, and not just $n$.}),
\beqa
M(z)&=&-\frac{16\pi G}{(n+1)\Omega_{n+1}}\,T^t{}_t\nn\\&=& m_0(z)+\ord{\frac1n}\,,
\eeqa
and the tension
\beqa\label{deftau}
\tau&=&-\frac{16\pi G}{\Omega_{n+1}}\,T^z{}_z\nn\\&=& m_0(z)-\frac{p_0^2(z)}{m_0(z)}+\partial_z p_0(z)+\ord{\frac1n}\,.\label{tension}
\eeqa

The tension $\tau$ is an intensive magnitude and in equilibrium configurations it must be uniform over the length of the string, \ie\ independent of $z$. This looks problematic, since $\tau$ in \eqref{tension} does appear to depend on $z$. However, when there is no dependence on time eqs.~\eqref{LOeqs} take the form
\beqa
p_0(z)&=&m_0'(z)\,,\\
0&=&\lp m_0''(z)+m_0(z)-\frac{\lp m_0'(z)\rp^2}{m_0(z)}\rp'\,.\label{steq0}
\eeqa
These imply that $\partial_z \tau=0$. Indeed, this condition is also verified at higher orders in $1/n$. In our choice of units, for a UBS (of any length) the constant is $\tau=1$ to all orders in $1/n$.

Note also that static solutions have zero momentum, $T_{tz}=0$. This also happens, as expected, to all higher orders.

The horizon of the NUBS is at
\beq
\sR=\sR_h(z)=m_0(z)+\ord{\frac1n}\,.
\eeq
Bear in mind that the actual area-radius is 
\beq\label{rh}
r_h=\sR_h^{1/n}=1+\frac{\ln m_0}{n}+\ord{n^{-2}}\,,
\eeq
which shows that in the large-$D$ approach the size of fluctuations of the horizon radius is $\ord{1/n}$.

If $r_h(z)$ varies along the string between $r_h^{\rm min}$ and $r_h^{\rm max}$, a useful measure of the non-uniformity is \cite{Gubser:2001ac}
\beq\label{lambd}
\lambda=\frac12\lp \frac{r_h^{\rm max}}{r_h^{\rm min}}-1\rp\,.
\eeq
Eq.~\eqref{rh} implies that $\lambda\sim 1/n$. Despite this limitation, in our study we will try to obtain large non-uniformities, approaching $O(1)$, by expanding to high orders in $1/n$.

We define a rescaled horizon entropy density, proportional to the area, as
\beqa
S(z)&=&\frac{\sqrt{n}}{\Omega_{n+1}}\textrm{Area}(z,\sR_h)\nn\\
&=&\sqrt{n G(z,\sR_h)}\,\sR_h^{\frac{n+1}{n}}(z)
= m_0(z)+\ord{\frac1n}\,.
\eeqa

The densities $M(z)$ and $S(z)$ can be integrated over the length of the string $L$ to obtain the total mass and entropy,
\beq\label{calma}
\mathbf{M}=\frac{L_{GL}^n}{L^{n+1}}\int_{-L/2}^{L/2} dz\, M(z)\,,\qquad \mathbf{S}=\frac{L_{GL}^{n+1}}{L^{n+2}}\int_{-L/2}^{L/2} dz\, S(z)\,,
\eeq
where
\beq
L_{GL}=\frac{2\pi}{k_{GL}}
\eeq
is the length of the uniform black string (of unit horizon radius) at the threshold of the GL instability. We have rescaled $\mathbf{M}$ and $\mathbf{S}$ by factors of $L^{-(n+1)}$ and $L^{-(n+2)}$, respectively, so as to render them invariant under changes of units. In other words, instead of having the units fixed by setting (as above) the horizon radius of the UBS equal to one, now the units are more sensibly set by the length of the compact circle $L$. In addition, we have introduced factors of powers of $L_{GL}$ so that the non-uniform branches start out at $\mathbf{M}=1$, $\mathbf{S}=1$ at all $n$, since it is practical to normalize quantities so that their value for the UBS at the GL threshold is equal to one at all $n$.


A convenient measure of the tension that is invariant under changes of units is the so-called ``relative binding energy'' $\mathbf{n}$ (or tension per unit mass and length) introduced in \cite{Harmark:2003dg,Harmark:2003eg},
\beq
\mathbf{n}=\lp\frac{L_{GL}}{L}\rp^n\frac{\tau}{\mathbf{M}}\,.
\eeq
Since the tension $\tau$ scales as (length)$^{n}$, the prefactor $L^{-n}$ makes the relative binding energy $\mathbf{n}$ scale invariant, while $L_{GL}^{n}$ normalizes it so that for the UBS at the GL threshold we have $\mathbf{n}=1$  at all $n$.

Finally, it is straightforward to compute the surface gravity at the horizon, $\kappa$, from which we define a rescaled surface gravity
\beqa\label{hkappa}
\hat\kappa&=&\frac{2}{n}\kappa\nn\\
&=& 1+\frac1n\lp\frac12\lp\frac{m'(z)}{m(z)}\rp^2-\frac{m''(z)}{m(z)}-\ln m(z)\rp+\ord{n^{-2}}\,.
\eeqa
The zeroth law of black holes requires that when the static equations of motion are satisfied $\kappa$ must be uniform over the length of the non-uniform black string. Since \eqref{deftau} and \eqref{hkappa} satisfy
\beq
\partial_z \kappa=-\frac1{2m(z)}\partial_z \tau+\ord{\frac1{n}}
\eeq
and we have seen that $\partial_z \tau=0$ then the zeroth law is indeed verified at LO. It also holds at higher orders. For all UBS (of any length), the constant value is $\hat\kappa=1$ at all $n$.

The surface gravity is of course proportional to the temperature, and they both scale like an inverse length. We will find convenient to employ a scale-invariant measure of the inverse temperature, which we take to be
\beq\label{boldbeta}
\bm{\beta}=\frac{L_{GL}}{L}\,\hat\kappa^{-1}\,.
\eeq
Again it is normalized so that $\bm{\beta}=1$ for a UBS at the GL threshold.

Summarizing, the boldfaced quantities $\mathbf{M}$, $\mathbf{S}$, $\mathbf{n}$ and $\bm{\beta}$ are the mass, entropy, relative binding energy, and inverse temperature for the NUBS in a unit circle $z\sim z + 1$, normalized relative to the values for the UBS at the GL threshold in that circle. Bearing in mind that the direction $z$ has been rescaled by a factor $\sqrt{n}$, we recover the physical \textsc{Mass}, \textsc{Entropy}, \textsc{Tension} and \textsc{Temperature} of the NUBS on a circle of proper physical $\textsc{Length}$  as
\beq
\textsc{Mass}=\mathbf{M}\,\frac{(n+1)\Omega_{n+1}}{16\pi G}\lp\frac{\sqrt{n}\,k_{GL}\,\textsc{Length}}{2\pi}\rp^{n}\textsc{Length}\,,
\eeq
\beq
\textsc{Entropy}=\mathbf{S}\,\frac{\Omega_{n+1}}{4G}\lp\frac{\sqrt{n}\,k_{GL}\,\textsc{Length}}{2\pi}\rp^{n+1}\textsc{Length}\,\,,
\eeq
\beq
\textsc{Tension}= \mathbf{n}\,\frac{\textsc{Mass}}{(n+1)\,\textsc{Length}}\,,
\eeq
\beq
\textsc{Temperature}=\frac{\sqrt{n}}{2k_{GL}\,\textsc{Length}}\bm{\beta}^{-1}\,,
\eeq
with $k_{GL}$ given by \eqref{kGLNNLO} (or more accurately \eqref{kGL4NLO}). The non-uniformity parameter $\lambda$ does not need any conversion to proper physical values.

\section{Perturbative Static Solutions and Static Critical Dimension}\label{sec:pertstatic}

Our aim now is to construct non-linear NUBS solutions and study their thermodynamic and  stability properties. In this section and in the next one we shall do this analytically in an expansion for small non-uniformity. 

\subsection{Perturbative NUBS}

We extend the linear perturbation analysis \eqref{linpert} of the static solution with $\Omega=0$, $k=k_{GL}$, to higher non-linear order in the amplitude of the perturbation. To this effect, we expand the collective variables as Fourier series of the form
\begin{equation}\label{epsexp}
 m(z) = 1+\sum_{j=1}^\infty \mu_j(\epsilon) \epsilon^j \cos\left(jk(\epsilon)z\right),\qquad  p(z) = \sum_{j=1}^\infty \nu_j(\epsilon) \epsilon^j \sin\left(jk(\epsilon)z\right)\,,
\end{equation}
where we allow the length of the compact circle,
\beq
L(\epsilon)=\frac{2\pi}{k(\epsilon)}\,,
\eeq
to vary with the non-uniformity perturbation parameter $\epsilon$. This is necessary since, purely for calculational simplicity, we are arbitrarily fixing a length scale by setting the constant, $z$-independent Fourier mode in $m(z)$ to be $1$ independently of $\epsilon$. The boldfaced quantities in the previous section are insensitive to this choice.

At the lowest, linear order in $\epsilon$, only the threshold static zero-mode, $j=1$, is present. Then, at each higher order in $\epsilon$ a new higher harmonic enters with $j=2,3,\dots$. At the same time, $k$ is modified as $\epsilon$ grows, \ie\ the periodicity is corrected. Therefore $\epsilon$ can be regarded as a mode-counting parameter. Bear in mind that besides the $\epsilon$ expansion, the functions $m$ and $p$ are also expanded in $1/n$. 
The calculations simplify slightly if we define $\epsilon$ so that 
\beq\label{firstmode}
\mu_1 = 1
\eeq
for all values of $n$. 

In order to illustrate the construction we solve the first few orders in the $\epsilon$ expansion for static solutions to leading order in $1/n$.
Plugging the ansatz \eqref{epsexp} in \eqref{steq0}\footnote{The second equation admits an obvious first integral, but this is not of much help for solving perturbatively the equations since the integration constant---the tension $\tau$---is $\epsilon$-dependent.} and successively solving the equations up to order $\epsilon^3$ we obtain
\beqa
\mu_2&=&\frac16+\ord{\epsilon^2}\,,\qquad \mu_3=\frac1{96}+\ord{\epsilon^2}\,,\\
k&=&1-\frac{\epsilon^2}{24}+\ord{\epsilon^4}\label{kperteps}\,.
\eeqa
Thus we obtain a static non-uniform black string solution with
\beqa
m_0(z)&=&1+\epsilon \cos k z+\frac{\epsilon^2}{6}\cos 2 k z +\frac{\epsilon^3}{96}\cos 3 kz +\ord{\epsilon^4}\,,\label{statpertm}\\
p_0(z)&=&-\epsilon\lp1-\frac{\epsilon^2}{24}\rp \sin k z-\frac{\epsilon^2}{3}\sin 2 k z -\frac{\epsilon^3}{32}\sin 3 kz +\ord{\epsilon^4}\,.\label{statpertp}
\eeqa

It is straightforward to extend these calculations to higher orders in $\epsilon$. At the same time, using the higher order equations we can also include corrections in $1/n$. In appendix~\ref{app:pertsol} we give the expansions of the Fourier coefficients $\mu_j(\epsilon)$ and $\nu_j(\epsilon)$, and the wavenumber $k(\epsilon)$, in powers of $\epsilon$ up to $\epsilon^{6}$, and in powers of $1/n$ up to $1/n^2$. \footnote{We have found evidence that the power series in $\epsilon$ may have finite convergence radius. The ratio between successive values of the coefficients $\mu_i$ appears to be close to $1/2$, which would imply that the applicability of the expansion is limited to $\epsilon<\sqrt{2}$.}
 
For solutions to leading order in $1/n$ we can immediately obtain the mass and entropy as\footnote{Note that this expansion seems to require $\epsilon^2\ll 1/n$, which originates in the factors $\sim (L_{GL}/L)^n$ in \eqref{calma}. This is not problematic, but instead we could equally well work with, \eg\ the mass-length $\mathbf{M}^{1/(n+1)}$ and entropy-length $\mathbf{S}^{1/(n+2)}$, which are equal to $1-\epsilon^2/24+\dots$ and only require $\epsilon\ll 1$.}
\beq\label{calmaLO}
\mathbf{M}=\mathbf{S}=1-\frac{n\epsilon^2}{24}+\ord{\epsilon^4}\,,
\eeq
and the tension as
\beq\label{tauLO}
\tau=1-\frac{\epsilon^2}{2}+\ord{\epsilon^4}\,.
\eeq
With our choice of units, the rescaled surface gravity \eqref{hkappa} is $\hat\kappa=1$ to leading order in $1/n$ independently of the deformation.

The parameter $\epsilon$ is not a very physical measure of the non-uniformity. Instead we can employ $\lambda$ in \eqref{lambd}, which for our solution above is
\beq\label{lambdaLO}
\lambda=\frac12\lp \lp\frac{m_0(0)}{m_0(\pi/k)}\rp^{1/n}-1\rp =\frac1{n}\lp\epsilon+\frac{17}{92}\epsilon^3+\ord{\epsilon^4}\rp\,.
\eeq

The results \eqref{calmaLO}, \eqref{tauLO}, \eqref{lambdaLO}  completely characterize weakly non-uniform black strings to leading order at large-$n$. At this order the mass and entropy of a given NUBS are exactly equal, the temperature is independent of the deformation, and therefore the thermodynamics of NUBS is rather uninformative. However, the inclusion of $1/n$ corrections yields much more interesting results. 

\subsection{Static critical dimensions}

With our definitions, a UBS with $\mathbf{M}=1$ is at the threshold of the GL instability, while a UBS with  $\mathbf{M}<1$ is unstable, and one with $\mathbf{M}>1$ is stable. 
Eq.~\eqref{calmaLO} says that, to leading order at large $n$, weakly-non-uniform strings have $\mathbf{M}<1$. Therefore, at sufficiently large $n$, for every weakly-unstable UBS there exists, nearby in solution space, a NUBS of the same mass. It is then possible that the UBS continuously evolves into a NUBS, in a smooth, second order transition between phases.

\begin{figure}[t]
\begin{center}
\includegraphics[width=.9\textwidth]{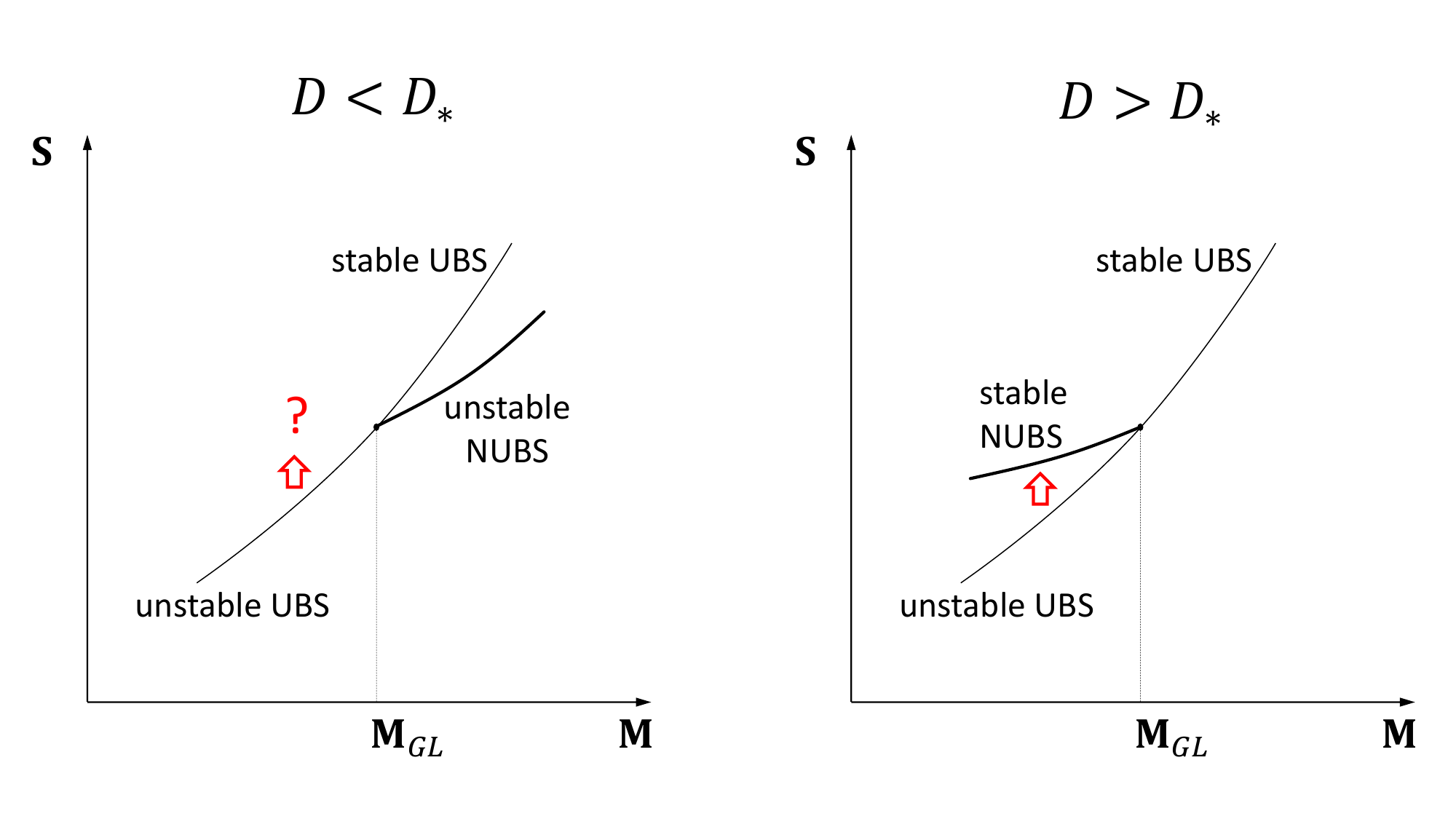}
\caption{\small Eq.~\eqref{MwnubsNLO}, and more accurately \eqref{Mwnubs}, shows that for $D<D_*$ there are no weakly non-uniform black strings with mass $\mathbf{M}<\mathbf{M}_{GL}=1$ in the range of unstable uniform black strings, so the latter cannot evolve smoothly into any of the former (the transition is then of first order into a phase further away, possibly other NUBS or BH). For $D>D_*$ the mass correction is reversed, and there do exist NUBS with $\mathbf{M}<\mathbf{M}_{GL}$ that unstable UBS can smoothly evolve into. The calculation of the entropy in \eqref{deltaA} shows that in $D>D_*$ this transition is consistent with the entropy law. Moreover, the analysis of sec.~\ref{sec:dyn} explicitly shows that weakly NUBS are dynamically linearly stable when $D>D_*$, and unstable when $D<D_*$.}\label{fig:Dstar}
\end{center}
\end{figure}

Finite $n$ effects can modify this behavior. 
Including the NLO terms the mass of a weakly-NUBS is
\beq\label{MwnubsNLO}
\mathbf{M}=1-\lp n-8\rp\frac{\epsilon^2}{24}+\ord{\epsilon^4}\,.
\eeq
Therefore, if $n<n_*=8$, \ie $D<D_*=12$, the mass is larger than 1: nearby a weakly-unstable UBS there is no NUBS of the same mass that it could continuously evolve into, neither by fluctuating in the microcanonical ensemble, nor through dynamical evolution in which (by axial symmetry) no energy is radiated. The UBS must then transit in a non-smooth, first order manner to another phase further separated in solution space. We illustrate the two situations in fig.~\ref{fig:Dstar}.

The value for the critical dimension can be improved by considering higher orders in $1/n$. For NUBS through 3NLO we obtain
\beq\label{Mwnubs}
\mathbf{M}=1-\frac{n\epsilon^2}{24}\lp 1-\frac{8}{n}-\frac{14}{n^2}+\frac{6-20\zeta(3)}{n^3}\rp+\ord{\epsilon^4}\,.
\eeq
Then, the change between $\mathbf{M}$ being smaller or larger than 1 at small non-uniformity occurs for
\beq\label{nstar}
n_*= 9.65\,,
\eeq
so that the smooth, continuous classical evolution of an unstable uniform black string to a weakly non-uniform one is only possible above the critical dimension
\beq\label{massDstar}
D_*= 13.65\,.
\eeq 
This agrees with remarkable accuracy with the numerical value $D_*\simeq 13.5$ obtained in \cite{Sorkin:2004qq}.\footnote{Already the leading order result $n_*=8$ provides a surprisingly good approximation. However, its calculation involves equating the LO and NLO in \eqref{Mwnubs}, which, strictly speaking, is not legitimate within perturbation theory. Nevertheless, these results for $n_*$ (and others closely related to it, as we will see) seem to stand up because the coefficient of the term $1/n$ gives a value quite larger than the correction from the term $1/n^2$, \ie\ because $8\gg 14/8$, and similarly at the next order. At present, all we can say is that this is a fortunate feature of the large-$n$ expansion.} In the large-$n$ expansion it had been obtained earlier in \cite{Suzuki:2015axa} (in a slightly different calculation to NNLO).

We must verify that the black hole entropy law allows to transit from the weakly-unstable UBS to a nearby NUBS of the same mass in $D>D_*$. This calculation requires a higher order perturbation: the first law of black holes implies that, for equal masses, the entropies of the UBS and the nearby NUBS are equal to order $\epsilon^2$ (in any $D$) \cite{Gubser:2001ac}\footnote{More precisely, the leading-order variations satisfy $\delta\mathbf{S}=\frac{n+1}{n}\delta\mathbf{M}$.}.  Moreover, to leading order in $1/n$ the entropy of all the solutions is the same as their mass \cite{Emparan:2015gva}. Therefore, in order to see the difference in the entropies we need the corrections at least at order $\epsilon^4/n$. Furthermore, revealing the reversal in the difference in entropies at the critical dimension requires at least one higher order in $1/n$. 

Bearing in mind that the entropy $\mathbf{S}$ of a UBS (not at the GL threshold!) with mass $\mathbf{M}$ is
\beq
\mathbf{S}_{\textrm{UBS}}=\mathbf{M}^{\frac{n+1}{n}}\,,
\eeq
we find that the relative entropy difference between NUBS and UBS of the same mass is
\beqa\label{deltaA}
\frac{\Delta \mathbf{S}}{\mathbf{S}}&=&\frac{\mathbf{S}}{\mathbf{S}_{\textrm{UBS}}}-1 \nn\\
&=&\lp 1-\frac{7}{n}-\frac{22}{n^2}-\frac{8(1+2\zeta(3))}{n^3}\rp \frac{\epsilon^4}{96n}+\ord{\epsilon^5}\,.
\eeqa
This changes sign at the critical dimension $n_*=9.59$, \ie
\beq\label{areaDstar}
D_*=13.59\,,
\eeq
in good agreement with \eqref{massDstar} to 3NLO accuracy. In fact, since it can be proven \cite{Gubser:2001ac,Sorkin:2004qq} that the first law of black holes implies (for any $n$) that
\beq\label{firstlaw}
\frac{\Delta \mathbf{S}}{\mathbf{S}}=\frac{n+1}{2n}\delta\mathbf{M}\lp \delta\bm{\beta}-\frac1n\delta\mathbf{M}\rp\,,
\eeq
where $\delta\bm{\beta}$ and $\delta\mathbf{M}$ are first-order variations (\ie here the corrections to $\ord{\epsilon^2}$),  then we see that $D_*$ must be the same whether we obtain it from $\delta\mathbf{M}=0$ or from $\Delta\mathbf{S}/\mathbf{S}=0$. Note, however, that the sign of the entropy difference is not determined by this equation.

Thus the large-$D$ expansion reproduces correctly all aspects of Sorkin's thermodynamic argument, including the precise value of $D_*$. As we will see in sec.~\ref{subsec:dyncrit}, the large-$D$ expansion also allows to directly establish that the smooth evolution of unstable UBS to stable NUBS is \textit{dynamically possible} above the critical dimension $D_*$.

\begin{figure}[t]
\begin{center}
\includegraphics[width=.9\textwidth]{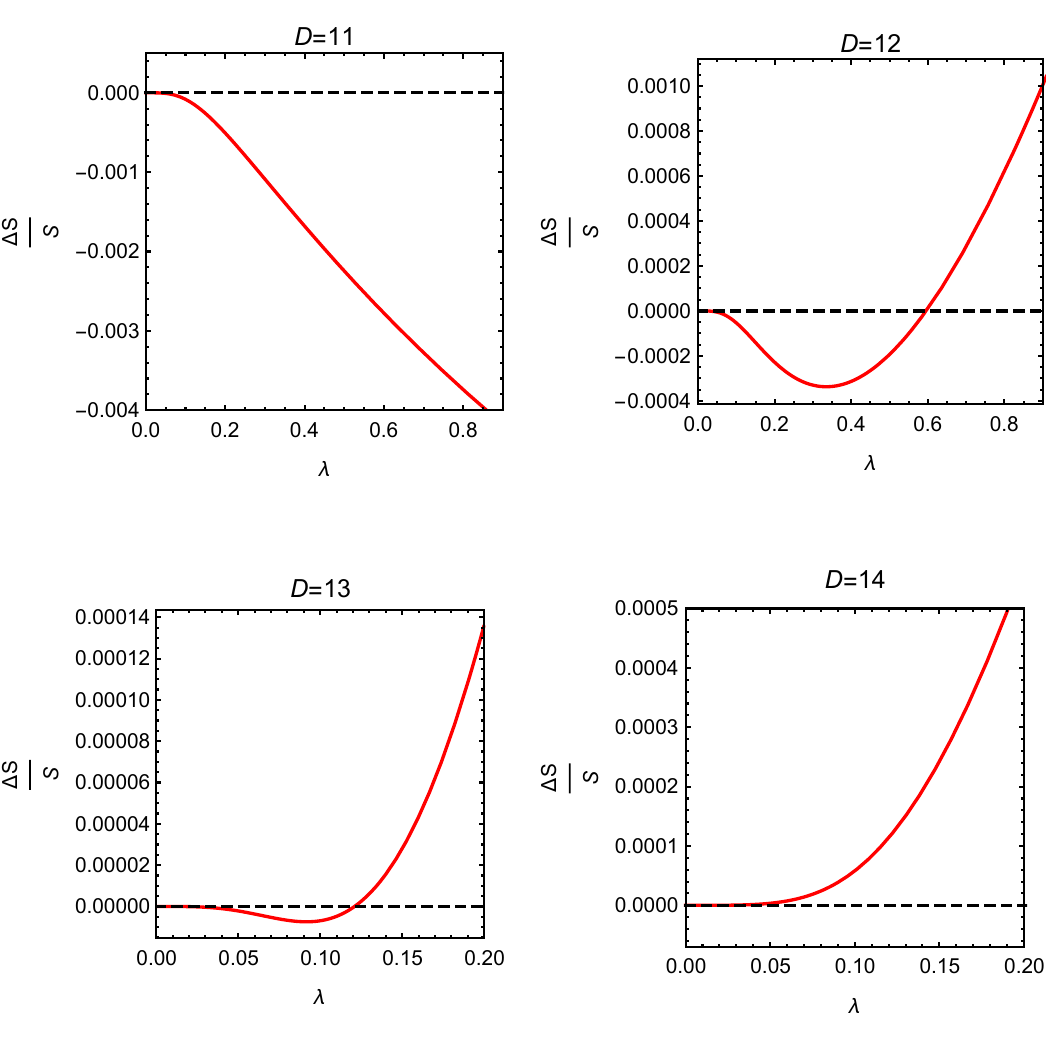}
\caption{\small Entropy difference \eqref{deltaA} between NUBS and UBS of the same mass as a function of non-uniformity $\lambda$ \eqref{lambd}. The NUBS in $D=12,13<D_*$ with $\Delta\mathbf{S}>0$ are thermodynamically stable even if they are below the critical dimension \cite{Figueras:2012xj}.
}\label{fig:DelA}
\end{center}
\end{figure}

We can further verify the presence of thermodynamically stable NUBS in $D=11,\ 12$, which are below the critical dimension, as found in \cite{Figueras:2012xj}. To this effect, in fig.~\ref{fig:DelA} we plot the entropy difference as a function of the deformation parameter $\lambda$. We observe that in $D=12,13<D_*$, there appear NUBS at finite deformation with positive entropy difference. The resemblance to fig.~7 of \cite{Figueras:2012xj} (even fairly quantitatively) is remarkable.

\subsection{Critical dimension for temperature}

The value we obtain for the inverse temperature
\beq\label{betanubs}
\bm{\beta}=1-\frac{\epsilon^2}{24}\lp 1-\frac8n-\frac2{n^2}-\frac{20\zeta(3)-6}{n^3}\rp +\ord{\epsilon^4,n^{-4}}
\eeq
is indeed such that \eqref{Mwnubs}, \eqref{deltaA}, \eqref{betanubs} do satisfy \eqref{firstlaw}.

Observe that there is a critical dimension
\beq
n_{*\beta}= 8.49\quad \ie\quad D_{*\beta}= 12.49
\eeq
at which the first order variation $\delta\bm{\beta}$ vanishes. If $D<D_{*\beta}$ then $\bm{\beta}$ increases as the non-uniformity appears, and it decreases if $D>D_{*\beta}$. 

Notice that $D_{*\beta}$ is different than $D_*$, as required by \eqref{firstlaw}. The numerical constructions of \cite{Figueras:2012xj} exhibit the same change in the behavior of $\bm{\beta}$ between $D=12$ and $D=13$ (see their fig.~4).

\subsection{4NLO corrections and non-convergence of the $1/n$ expansion}\label{subsec:nonconverg}

Ref.~\cite{Emparan:2015rva} argued that the large-$n$ expansion is not convergent, but only asymptotic. Non-perturbative effects that couple the near-horizon region to the far region turn out to limit the reliability of the perturbative expansion. The size of these effects is $\sim n^{-n/2}$, so they become comparable to perturbative corrections $1/n^k$ at order $k\sim n/2$. Therefore, when calculating the critical dimension $n_*\simeq 9$, we may expect that terms from $1/n^4$ or $1/n^5$, \ie\ 4NLO or 5NLO corrections, begin to show poorer convergence.

We have found evidence of this. We have managed to extend our results for weakly-non-uniform black strings up to 4NLO, which adds to \eqref{Mwnubs} the term
\beq
\delta^{(4)}\mathbf{M}=\frac{\epsilon^2}{18n^3}\lp 21+2\pi^4+30\zeta(3)\rp\,.
\eeq
Then the critical dimension for $\mathbf{M}$ is determined by requiring that
\beq
n_*=8\lp1+\frac{1.75}{n_*}+\frac{2.26}{n_*^2}+\frac{42.0}{n_*^3}\rp\,.
\eeq
The coefficient of the last term is unusually large, with the effect that it corrects $n_*$ by a larger amount than the previous term. More explicitly, at each successive order we get
\beq
n_*= 8~/~ 9.48~/~9.65~/~9.93\,,
\eeq
which signals a loss of convergence in the last correction. The latter, then, should not be trusted.

The same breakdown of the expansion is observed, at the same order, in the calculation of $n_{*\beta}$. At 4NLO we get
\beq
\delta^{(4)}\bm{\beta}=-\frac{24}{n}\delta^{(4)}\mathbf{M}\,,
\eeq
which yields
\beq
n_{*\beta}= 8~/~ 8.24~/~8.49~/~8.92\,.
\eeq
Again, we deem the last value unreliable.

These results may be taken as suggesting that it would not be useful to try to obtain higher orders in the expansion. However, bear in mind that this can depend on the specific quantity that is computed. For instance,  it was shown in \cite{Emparan:2015rva} that while the 4NLO corrections do not uniformly improve the values for the quasinormal frequencies of UBS, they nevertheless yield the best approximation to $k_{GL}$ even down to $D=6$, where the accuracy is within $2.4\%$. We will find in sec.~\ref{subsec:thermonubs} that 4NLO results also give excellent agreement for $\mathbf{M}$ and $\mathbf{S}$ above but also below the critical dimension. Perhaps in these cases one needs to go to 5NLO to find evidence of non-convergence in these dimensions---our readers are invited to try their hand at such calculations.

\section{Stability of NUBS and Dynamical Critical Dimension}\label{sec:dyn}

As far as we are aware, no dynamical study of the stability of non-uniform black strings has been performed yet in any finite number of dimensions.\footnote{The closest is the study in \cite{Figueras:2012xj} using local Penrose inequalities.} The large-$D$ expansion simplifies the task enormously, even allowing analytical investigation. 

\subsection{Quasinormal modes of NUBS}

Let us consider a static solution $m_s(z)$, $p_s(z)$. 
We perturb it by adding time-dependent terms
\begin{equation}
m(t,z) = m_s(z)+e^{\Omega t}\delta m(z),\qquad  p(t,z) = p_s(z)+e^{\Omega t}\delta p(z),
\end{equation}
and expand the equations to linear order in $\delta m$ and $\delta p$. If the initial solution is periodic over an interval of length $2\pi/k$, then the perturbation admits an expansion as Fourier series\footnote{We exclude $j=0$ since these are exact zero modes that can be absorbed in uniform rescalings and boosts.} 
\begin{equation}\begin{split}
\delta m(z) = \sum_{j=1}^\infty\delta m^{(+)}_j \cos(jk z)+\delta m^{(-)}_j \sin(jkz),\\
\delta p(z) = \sum_{j=1}^\infty\delta p^{(+)}_j \sin(jkz)+\delta p^{(-)}_j \cos(jkz).
\end{split}\end{equation}
We consider static solutions that have the symmetry $m_s(z)=m_s(-z)$, $p_s(z)=-p_s(-z)$, and then the spectrum can be split into even $(+)$ and odd $(-)$ modes.

Observe that $\Omega$ has dimensions of inverse length, so if we want to measure it, as we are doing for all other quantities, in units of the length $L$, then we must use the scale-invariant frequency
\beq
\bm{\Omega}=\frac{L}{L_{GL}}\Omega\,,
\eeq
and the corresponding physical frequency will be
\beq
\textsc{Omega}=\frac{\bm{\Omega}}{\sqrt{n}\, k_{GL} \textsc{Length}}\,.
\eeq

When we take the static solution  to be perturbative in $\epsilon$ and in $1/n$, the coefficients of the time-dependent fluctuation, $\delta m_j^{(\pm)}$ and $\delta p_j^{(\pm)}$, will also be power series of $\epsilon$ and $1/n$. 
At the lowest order in both expansions we are perturbing the uniform black string with $k=1+\ord{\epsilon,1/n}$. It is clear from our earlier result \eqref{LOspectrum} that the modes with wavenumber $j k=j+\ord{\epsilon,1/n}$ have 
\beq
\bm{\Omega}=-j(j\pm1)+\ord{\epsilon,1/n}\,.
\eeq
Therefore perturbation modes of a weakly-NUBS (small $\epsilon$) with $j> 1$ will have $\Omega<0$ and so these perturbations are stable. This was of course expected, since these `overtones' all have wavelengths shorter than the `fundamental' threshold mode $j=1$. While it would be possible to compute corrections in $\epsilon$ and in $1/n$ to this result, there appears to be little motivation for it.

It then remains to study perturbations that are dominated by fundamental normal modes with $j=1$, which, at small $\epsilon$, lie near the stability threshold with $\bm{\Omega}=\ord{\epsilon}$. It suffices to focus on even modes (odd ones are simply a translation of them), \ie\ $\delta m(z)=\cos k z+\ord{\epsilon,1/n}$, $\delta p(z)=-\sin k z+\ord{\epsilon,1/n}$.

We begin working at leading order in $1/n$, and perturb the static solution we found in section~\ref{sec:pertstatic}, with $m_s(z)$ and $p_s(z)$ given by \eqref{statpertm} and \eqref{statpertp}, and $k$ by \eqref{kperteps}. Solving for the coefficients $\delta m_j^{(\pm)}$, $\delta p_j^{(\pm)}$ perturbatively in $\epsilon$ up to cubic order we obtain
\beqa
\delta m(z)&=&\cos k z +\epsilon\,\frac13 \cos 2 k z+\epsilon^2\, \frac1{32}\cos 3k z + \ord{\epsilon^3}\,,\\
\delta p(z)&=&-\lp 1-\frac{\epsilon^2}{8}\rp\sin k z -\epsilon\,\frac23 \sin 2 k z-\epsilon^2\, \frac3{32}\sin 3k z + \ord{\epsilon^3}
\eeqa
with
\beq
\bm{\Omega}=-\frac{\epsilon^2}{12}+ \ord{\epsilon^3}\,.
\eeq
Since $\bm{\Omega}<0$, we conclude that, to this order in small $\epsilon$, and to leading order for $n\to\infty$, weakly non-uniform black strings are stable. This LO calculation can be readily carried over to two higher orders in $\epsilon$, where we find 
\begin{equation}\label{Omeps}
\bm{\Omega}(\epsilon)
   =-\frac{\epsilon^2}{12}\left(1+\frac{7\epsilon^2}{16}+\frac{75497\epsilon^4}{414720}\right)+\ord{\epsilon^8},
\end{equation}
so $\bm{\Omega}(\epsilon)<0 $ persists to this order. This analytical argument for the stability of NUBS when $D\to\infty$ is in perfect agreement with their numerically observed stability under dynamical evolution of the LO equations \eqref{LOeqs} \cite{Emparan:2015gva}.

\subsection{Dynamical critical dimension}\label{subsec:dyncrit}

It becomes even more interesting when we add the first correction for finite $n$. In this case we obtain
\beq\label{Omegae21n}
\bm{\Omega}=-\frac{\epsilon^2}{12}\lp 1-\frac{10}{n}\rp + \ord{\epsilon^4}\,.
\eeq
Here we see the appearance of a ``dynamical critical dimension'', $n_*=10$, \ie\ $D_*=14$, such that for $n<n_*$ weakly non-uniform black strings are dynamically unstable, while for $n>n_*$ they are stable. 

This critical dimension is very close to the one we found from the thermodynamical analysis.
The agreement improves with the 3NLO result
\begin{equation}\label{Omeps1n}
\bm{\Omega}
   =-\frac{\epsilon^2}{12}\left(1-\frac{10}{n}+\frac{6-2\zeta(2)}{n^2}-\frac{6-4\pi^2+20\zeta(3)}{n^3}\right)+\ord{\epsilon^4},
\end{equation}
so the critical value where $\bm{\Omega}$ changes sign is corrected to
\begin{equation}
n_* =9.62\,,
\eeq
\ie\
\beq\label{dynDstar}
D_*=13.62\,.
\end{equation}
Now this is the same result (well within the expected accuracy) as  obtained in \eqref{massDstar} and \eqref{areaDstar} from the thermodynamics of the phase space of \textit{static} solutions. Going one order higher the expansion appears not to converge, but this might be expected from our discussion in sec.~\ref{subsec:nonconverg}.\footnote{At 4NLO we find $\delta^{(4)}\bm{\Omega}=\epsilon^2(8+5\pi^2+49\pi^4/45)/(6 n^4)$, which would yield $n_*=9.96$.}

The connection between the thermodynamic critical dimension of NUBS and the change in their dynamical quasinormal stability was expected on general grounds, but so far it had not been verified explicitly. The $1/D$ expansion has allowed us to establish it with excellent accuracy. 

\subsection{Quasinormal stability of NUBS and Poincar\'e turning points}

We have extended the calculation of the lowest quasinormal frequencies of NUBS to higher orders in the non-uniformity, see \eqref{Omegafund}. The most salient aspect of the result is to show that large enough non-uniformity can change the stability of NUBS in some dimensions below the critical value.

In order to illustrate this phenomenon, let us keep only the next-to-leading-order terms in both $\epsilon$ and in $1/n$, and write them as
\begin{equation}\label{Omeps1n}
\bm{\Omega}(\epsilon)
   =-\frac{\epsilon^2}{12}\left(1+\frac{7\epsilon^2}{16}\right)
   \left(1-\frac{10}{n}\lp1-\frac{3\epsilon^2}{20}\rp\right)\,.
\end{equation}
%
%
To leading order in the non-uniformity this is the same as \eqref{Omegae21n}, which showed that NUBS in $n<n_*=10$ are unstable. However, when the next non-uniformity order, $\epsilon^4$, is included, the instability gets weaker. More precisely, a NUBS with $\epsilon=\epsilon_0$, where
\beq\label{criteps}
\epsilon_0=\sqrt{\frac{2\lp n_*-n\rp}{3}}\simeq 0.82\sqrt{n_*-n}\,,
\eeq
has a zero mode instead of a negative mode, and a NUBS with $\epsilon>\epsilon_0$ would be dynamically linearly stable even if it is below the critical dimension. 

This finding ties in very well with the presence of a turning point at finite non-uniformity in the mass of NUBS in $n<n_*$.\footnote{This is related to, but not the same as the existence of stable NUBS below the critical dimension, discussed in \cite{Figueras:2012xj} and in sec.~\ref{subsec:thermonubs}.} To see the relation clearly, let us write the derivative of the mass with respect to non-uniformity as
\beq
\frac{\partial\mathbf{M}}{\partial\epsilon^2}=-\frac{n\mathbf{M}}{24}\lp1+\frac{83}{144}\epsilon^2\rp\lp 1-\frac{8}{n}\lp 1-\frac{23}{72}\epsilon^2\rp\rp\,.
\eeq
The last term in brackets reveals that, below the (at this order, in this case) critical dimension $n_*=8$, the mass reaches a turning point when the non-uniformity is $\epsilon=\epsilon_{tp}$, with
\beq\label{criteps2}
\epsilon_{tp}=\sqrt{\frac{9\lp n_*-n\rp}{23}}\simeq 0.63\sqrt{n_*-n}\,.
\eeq
This result is very close to $\epsilon_0$. Indeed, the two results are expected to coincide: Poincar\'e's turning-point method says that a solution at a turning point in phase space must have one zero mode (for at least one kind of perturbation). Although $\epsilon_0$ and $\epsilon_{tp}$ are not exactly the same, they are sufficiently close, within the approximations we have made, to validate the conclusion that they conform to this argument.\footnote{ In order to make this agreement more precise we would need better accuracy in the non-uniformity than we have obtained, even more so if we are interested in turning points at integer values of $n<n_*$, which lie far from the GL point. Calculations to higher non-uniformity than in \eqref{deltaA} would also be needed to verify that around the turning point the solution with $\bm{\Omega}<0$ has larger entropy than the solution with the same mass and $\bm{\Omega}>0$, as required by Poincar\'e's method.}

So, once again, we see that the large $D$ expansion is an efficient means of establishing the links between local thermodynamical stability and linear dynamical stability that exist in this system.

\section{Highly non-uniform black strings}\label{sec:nums}

In addition to the analytic expansions for small non-uniformity, we have explored numerically larger non-uniformity in strongly non-linear regimes. We have done this by finding highly deformed static NUBS, and by time evolution of UBS deep into a non-linear regime. 

The effective large-$D$ equations can be written and solved in two different ways. The first one is as a \textit{sequential} set of $j+1$ equations, one equation for each term $m_j(t,z)$ of the pertubative expansion of $m(t,z)$, \eqref{nested}. The equation for $m_j(t,z)$ involves the solutions at lower orders, $m_i(t,z),\,i=0,\dots,j-1$.\footnote{If the equations were linear, the lower order solutions would yield sources for  $m_{j} (t,z)$. Here the equations are non-linear, so $m_{i<j} (t,z)$ also appear as coefficients.} Crucially, $n$ does not appear anywhere in these equations: it only enters, as a free parameter, when we recombine the solutions $m_j(t,z)$ into a finite series in $1/D$ to recover the solution $m(t,z)$ up to $j$NLO. Thus, in this approach $n$ remains a continuous, analytically tractable parameter, not only in the equations but also in the solutions.

The second approach consists of solving an \textit{inclusive} equation for the total variable $m(t,z)$, \eqref{nested}, which includes at once all the corrections up to a given order. That is, to any order in the $1/n$ expansion we solve only one equation for $m(t,z)$, instead of $j+1$ equations for $m_j(t,z)$ in the sequential approach. The price to pay is that this single equation now involves $n$ explicitly and therefore, if we want to integrate it numerically, we must assign a specific value to $n$. That is, in contrast to the sequential approach, the integration yields a solution for a specific value of $n$, and so we must solve the whole equation anew to obtain the solution for another value of $n$.

Both approaches should yield compatible results within a given order of the expansion, but we have found each one preferrable for a different problem. 

The sequential approach can easily be applied to the solution of the ODEs of the static system, and this allows to efficiently scan the phase space of static solutions, including the unstable phases that would not be visible in a dynamical evolution. Once we obtain solutions for $m_0(z)$, $m_1(z)$, $m_2(z)$, $m_3(z)$ and $m_4(z)$, the complete solution $m(z)$ is known to 4NLO for any value of $n$. The dimensionality of the space of parameters that one needs to scan numerically is then reduced by one. This approach has the drawback that for some values of $n$ the solutions are unphysical (for instance, with negative mass density or tension) and we must identify and remove them out of the space of static solutions.

The inclusive approach is more useful for the time evolution of the effective equations. We studied the endpoint of the dynamic evolution starting from a slightly perturbed unstable black string. We are able to verify stability beyond the linear analysis, but we find an apparent instability in a regime where the NUBS should presumably be stable, according to the static results.  Already at NLO, there is a $D$-dependent limiting thickness, $\mathbf{M}_{\textrm{lim}}(D)$, where the string becomes too inhomogeneous and the numerical code breaks down. This indicates the point at which the non-uniformity of the black string becomes so large that the effective equations of the $1/D$ expansion cease to be reliable. There is no true (finite-$D$) physics that corresponds to the phenomena that we observe there. Surprisingly, there seems to be a scaling behavior in the non-uniformity for the large-$D$ breakdown, as  $\mathbf{M}_{\textrm{lim}}(D) \sim D^{-D/4}$, similar to the one encountered for the merger point.

To solve the equations we decomposed the functions $m_j(z)$ as (truncated) Fourier series. The Fourier coefficients were then fitted by a Levenberg-Marquardt algorithm in order to satisfy the effective equations. Additionally, the same results were computed independently up to NLO using a Chebyshev grid, and a Newton-Raphson relaxation.

\subsection{Thermodynamic properties of NUBS branches}\label{subsec:thermonubs}

In our computations of static NUBS using the sequential approach we have chosen to terminate the branches, in each dimension, when the NUBS reach zero tension. Although this is an estimate and not an accurate determination of the actual merger transition to the BH phase, we will see that it is remarkably close to the endpoints of the branches obtained through full-numerical solution of the Einstein equations.

In \cite{Suzuki:2015axa} an analytical approximation to the profile of zero-tension solutions was made. Their shape was found to be fairly close to what one would expect for a BH. The NNLO horizon position in this case is
\beq\label{Rhzero}
\sR_h(z)\sim e^{-z^2/2}\lp 1-\frac{z^4}{n}+\frac{3z^8-16}{96n^2}\rp\,,
\eeq
which indicates that when the BH fills up the length of the compact circle, this length  will scale as $L\sim \Delta z\sim n^{1/4}$. From this one can readily estimate that the (static) merger transition between NUBS and BH occurs for values that scale as
\beq\label{zeroTscaling}
\mathbf{M}^{1/(n+1)},\,\bm{\beta}\sim n^{-1/4}\,,
\eeq
since both these quantities are defined to be inversely proportional to $L$ (see \eqref{calma} and \eqref{boldbeta}).

\begin{figure}[H]
\begin{center}
\includegraphics[height=200pt]{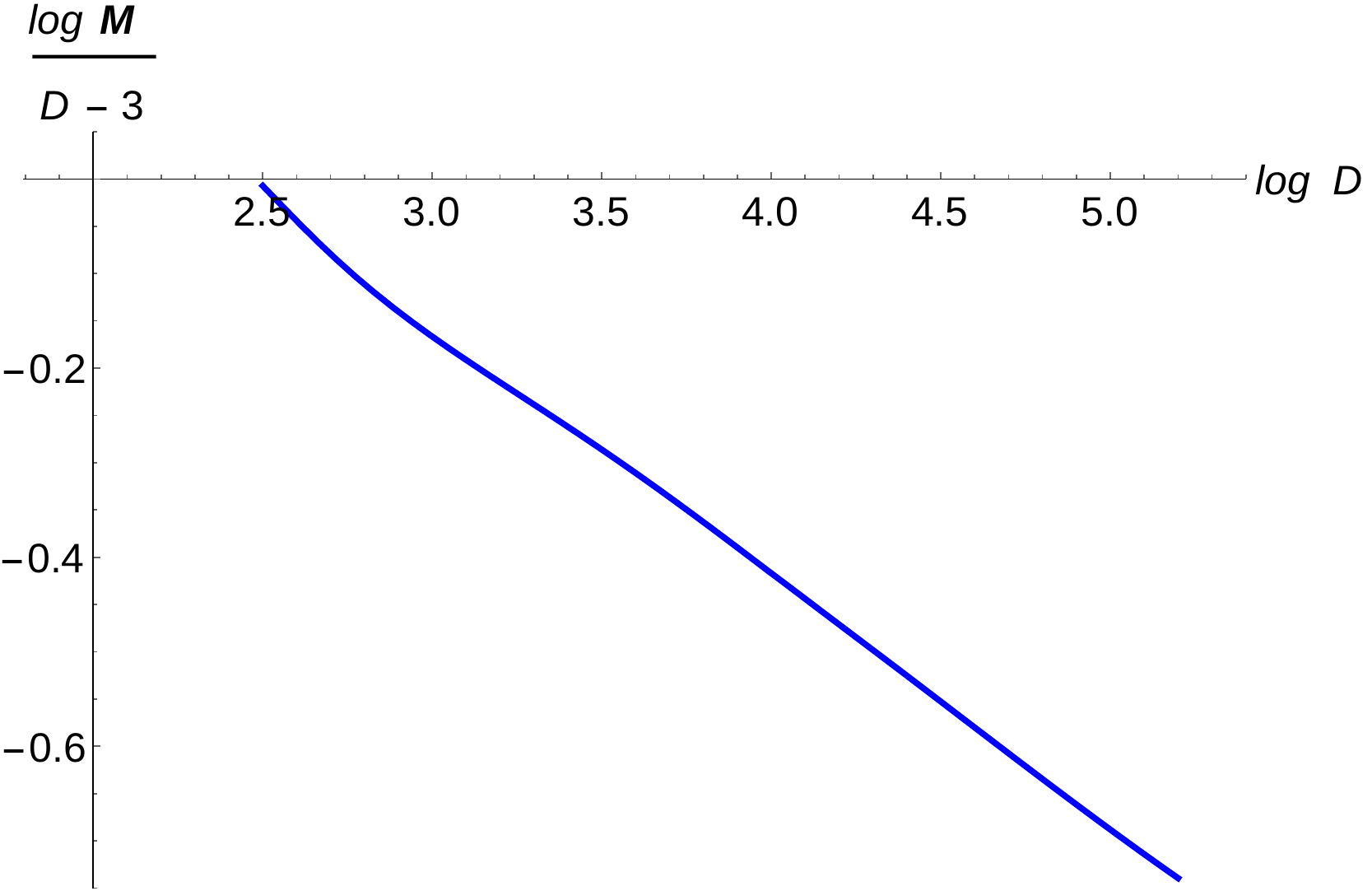}
\caption{\small Scaling behavior of the zero-tension curve $\mathbf{M}_{\textrm{min}}(D)$ that we take as a proxy for the merger transition between static NUBS and BH phases. The curve  (obtained at 4NLO) falls off as $D^{-1/4}$, in excellent agreement with the analytical estimate \eqref{zeroTscaling}.}\label{fig:LogLogPlot}
\end{center}
\end{figure}

Our numerical results clearly confirm this scaling behavior, exhibiting its onset already at dimensions as low as $n\approx 8$, \ie $D\approx 12$, see fig.~\ref{fig:LogLogPlot}. We denote the zero-tension values of the NUBS mass as
\beq\label{Mmin}
\mathbf{M}=\mathbf{M}_{\textrm{min}}(D)\sim D^{-D/4}\,.
\eeq
These are the curves shown in blue in figs.~\ref{fig:phases} and \ref{fig:LogLogPlot}. 

The following figures \ref{fig:BetaM}, \ref{fig:BetaA}, \ref{fig:BetaTension}, \ref{fig:LambdaBeta}, show our numerically computed branches of NUBS that are extended until they reach zero tension. We also present the results of our analytical calculations in the expansion in $\epsilon$ which, as we could expect, are accurate only as long we do not depart too far from the beginning of the branch. Furthermore, we include a comparison with the results obtained in \cite{Figueras:2012xj} in $D=13$ and $D=14$ (just below and above $D_*$) through full numerical solution of the Einstein equations. 
 
Figs.~\ref{fig:BetaM} and \ref{fig:BetaA} show the total mass and entropy of NUBS as functions of the inverse temperature. We are not presenting diagrams of $\mathbf{S}$ vs.\ $\mathbf{M}$ since (as is indeed apparent by comparing these diagrams) the difference between them is very small and the curves for NUBS are too close to the curves for UBS to give a useful image.

The figures show that our 4NLO results for the mass and entropy provide an excellent match to the calculations in \cite{Figueras:2012xj}. 
We emphasize that there is no free parameter in this comparison. The quantitative agreement is remarkable not only at the GL point but also further along the branches. This is strong evidence that the large-$D$ expansion, with higher order corrections included, can work well even for inhomogeneities of order one despite its apparent limitation to inhomogeneities $\sim 1/D$.  We are particularly surprised by how well the zero-tension condition for the termination of the branches appears to agree with the limits found in \cite{Figueras:2012xj}. This is presumably due to the fact, already observed in \cite{Suzuki:2015axa}, that the zero-tension solution \eqref{Rhzero} appears to capture well the geometry near the equator of a BH (but not near the axis where $\sR_h\to 0$). Then the solution \eqref{Rhzero} reproduces correctly the mass and area of the BH since at large-$D$ these quantities are dominated by their values near the equatorial bulge.

\begin{figure}[H]
\begin{center}
\includegraphics[width=200pt]{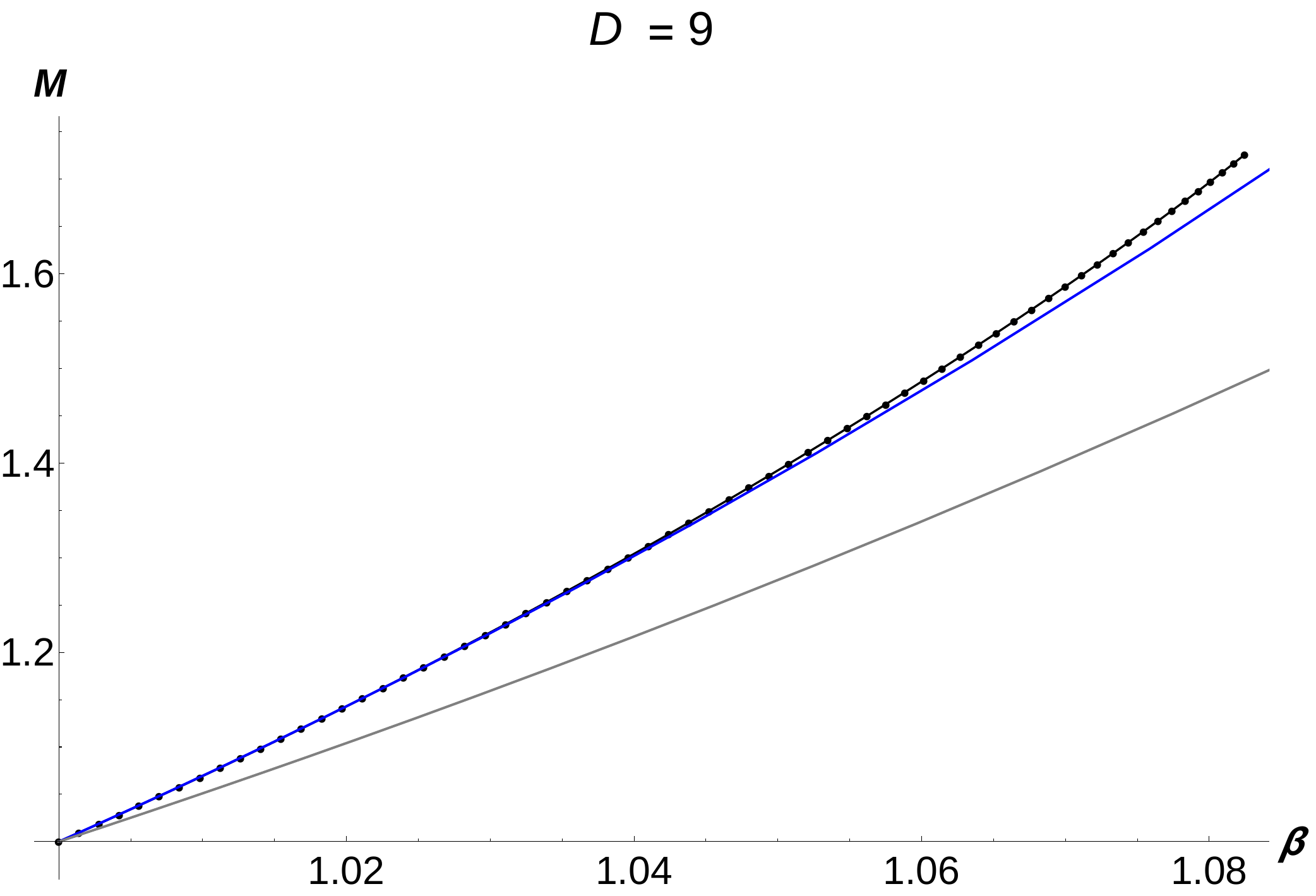}
\includegraphics[width=200pt]{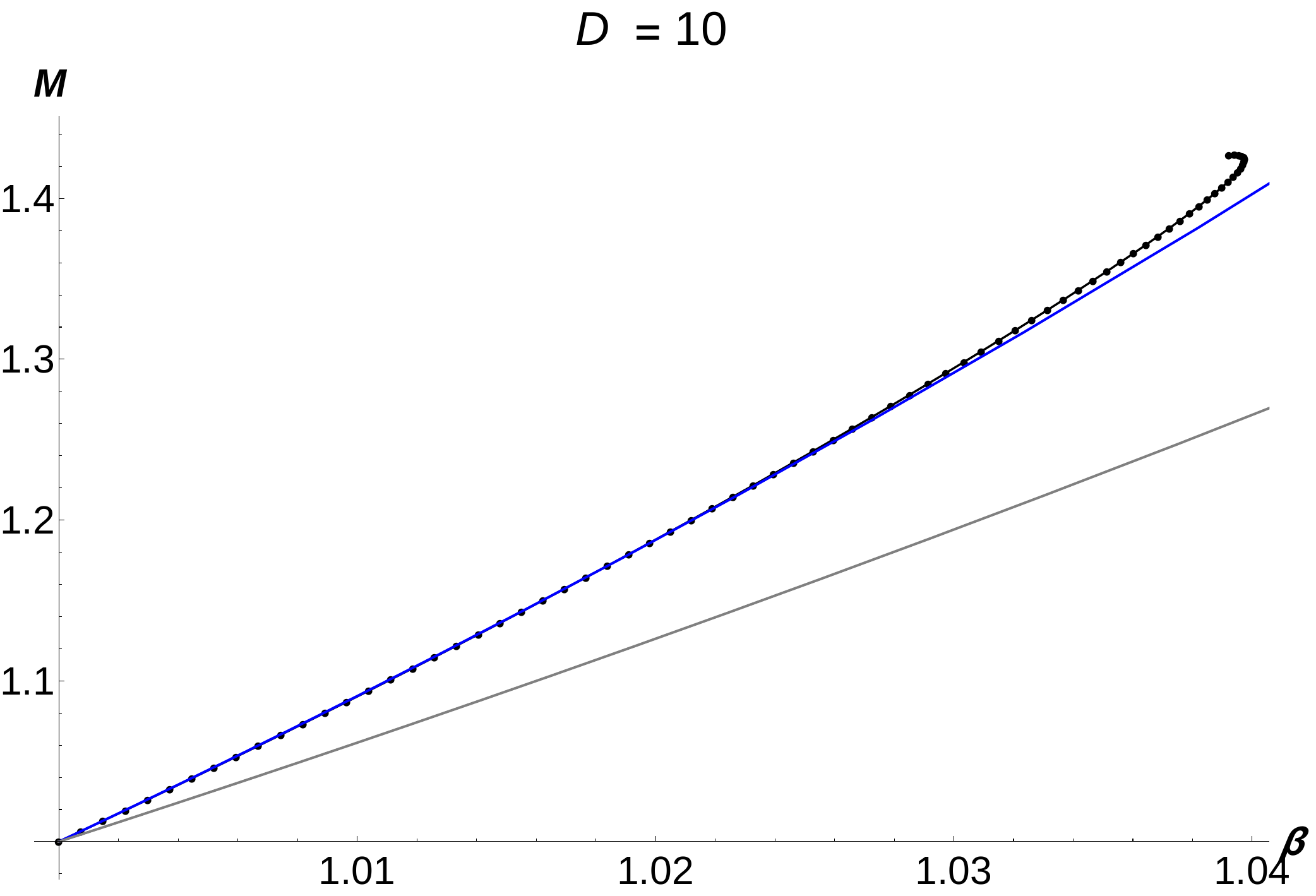}
\hfill\break
\hfill\break
\includegraphics[width=200pt]{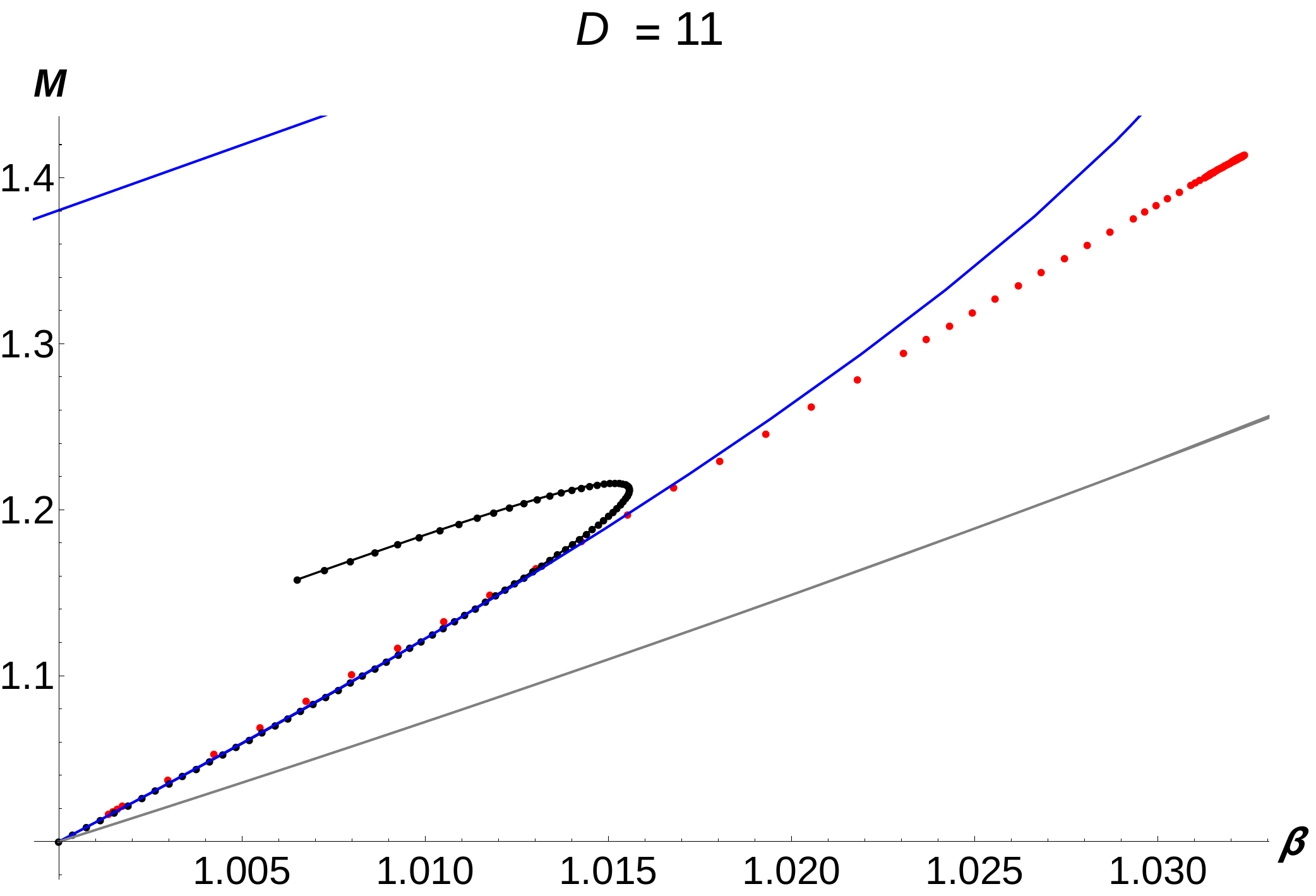}
\includegraphics[width=200pt]{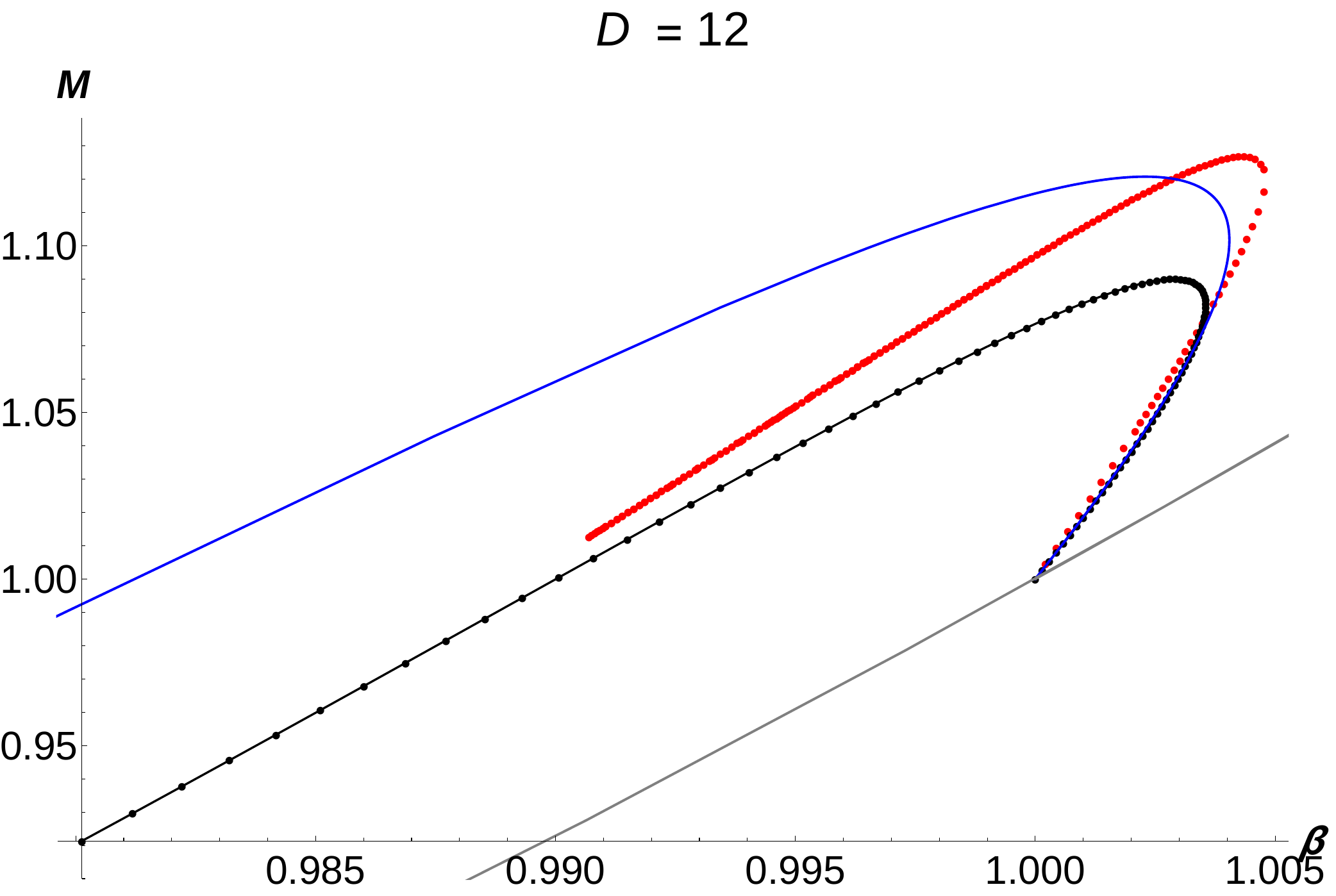}
\hfill\break
\hfill\break
\includegraphics[width=200pt]{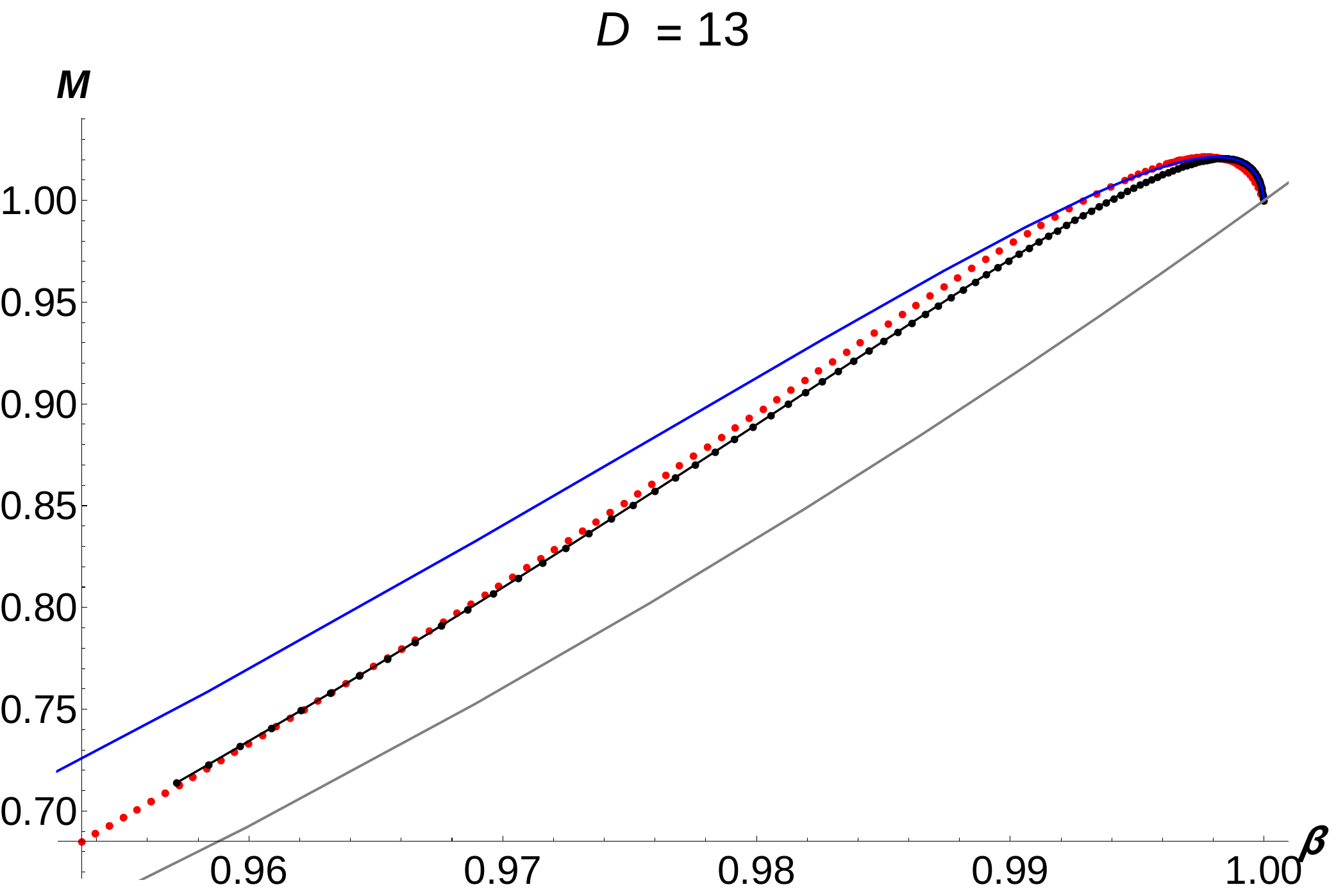}
\includegraphics[width=200pt]{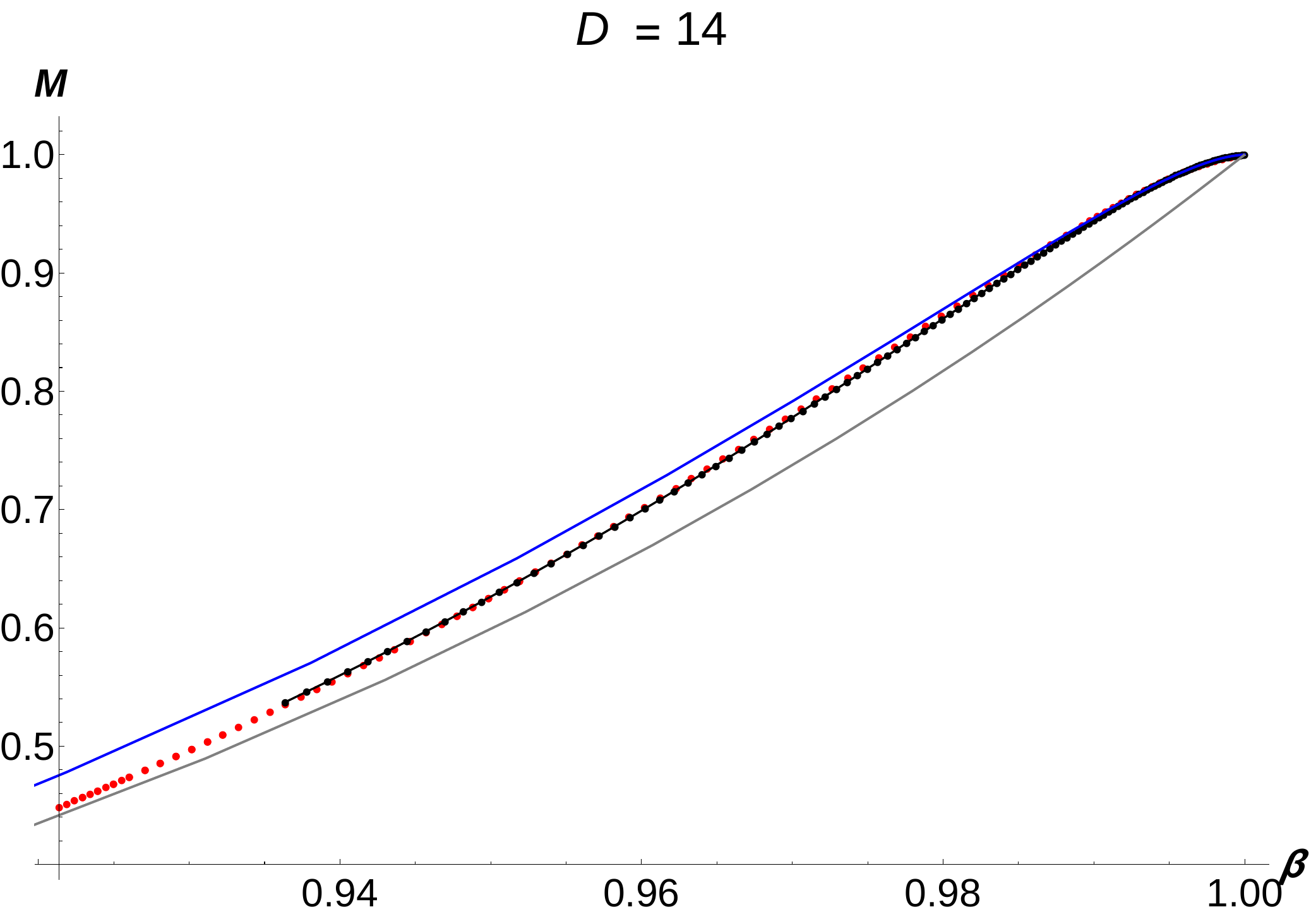}
\caption{\small Mass $\mathbf{M}$ vs.\ inverse temperature $\bm\beta$. Black dots: large-$D$ numerical results for NUBS.  Blue solid: large-$D$ perturbative solution for NUBS. Red dots: finite-$D$ full-numerical NUBS in \cite{Figueras:2012xj}. Gray solid: uniform black string ($\mathbf{M}_{\textrm{UBS}}= \bm{\beta}^n$). From the branching point at $\mathbf{M}=\bm{\beta}=1$, we see that $\bm{\beta}$ increases when $D<12.5$ and decreases when $D>12.5$, while $\mathbf{M}$ increases when $D<13.6$ and decreases when $D>13.6$.}\label{fig:BetaM}
\end{center}
\end{figure}

\begin{figure}[H]
\begin{center}
\includegraphics[width=200pt]{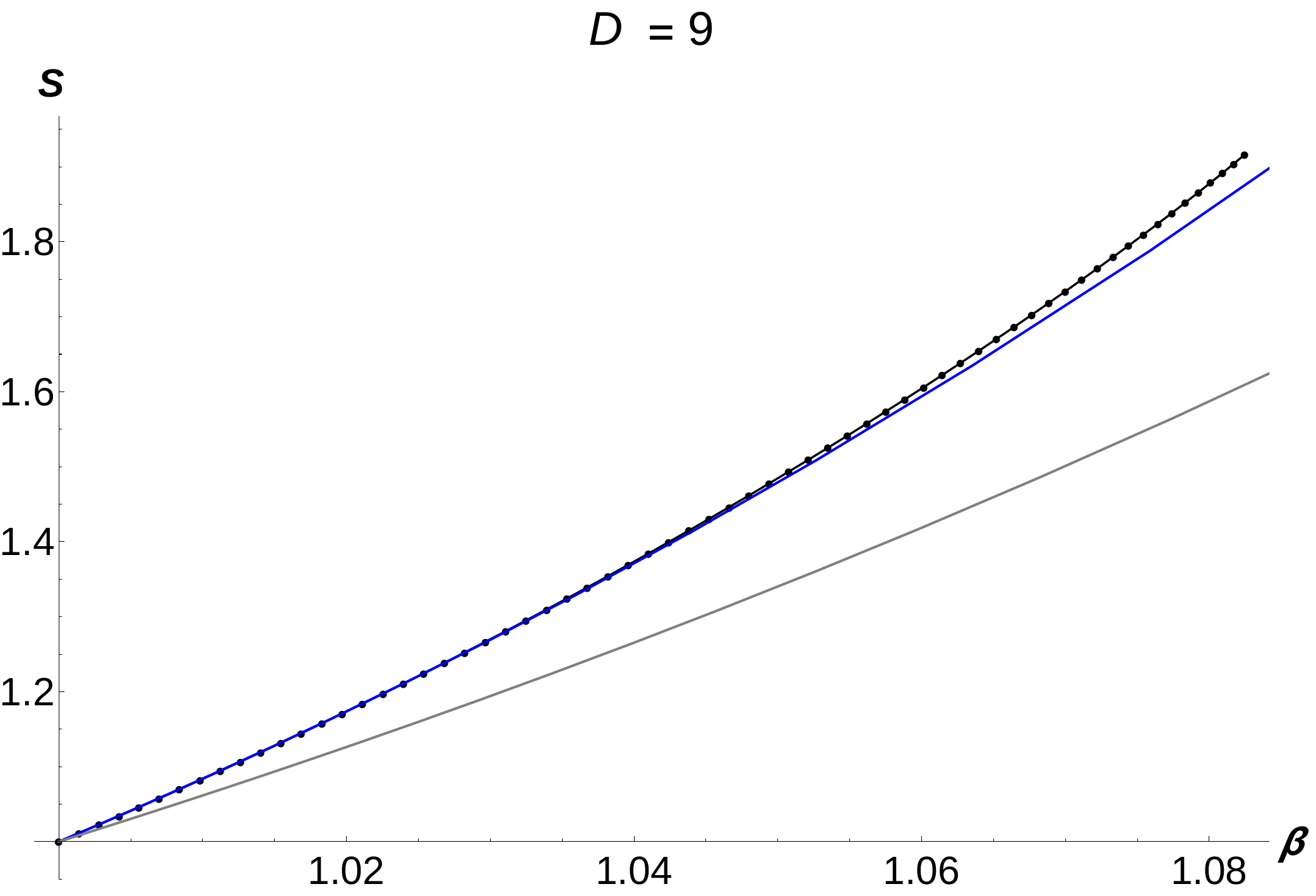}
\includegraphics[width=200pt]{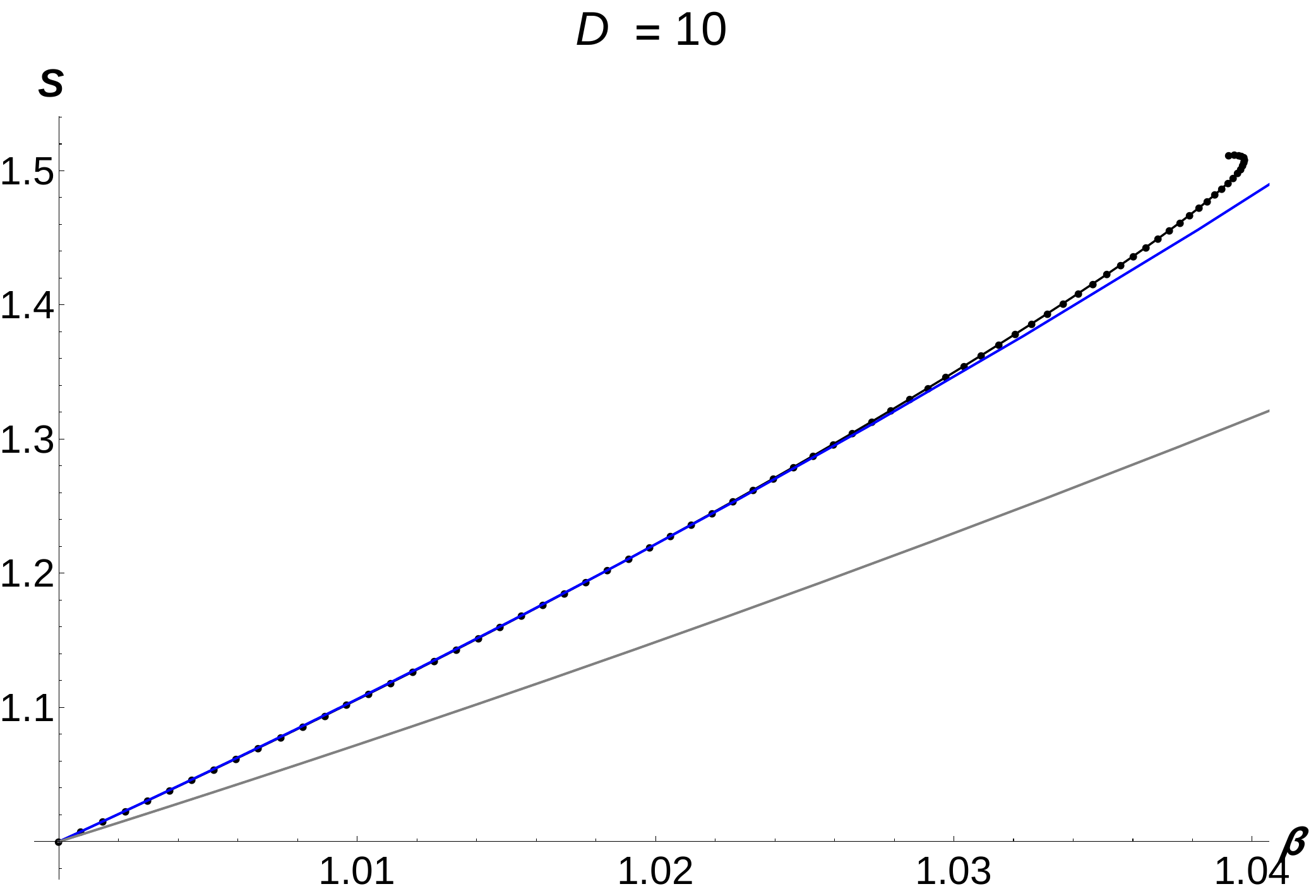}
\hfill\break
\hfill\break
\includegraphics[width=200pt]{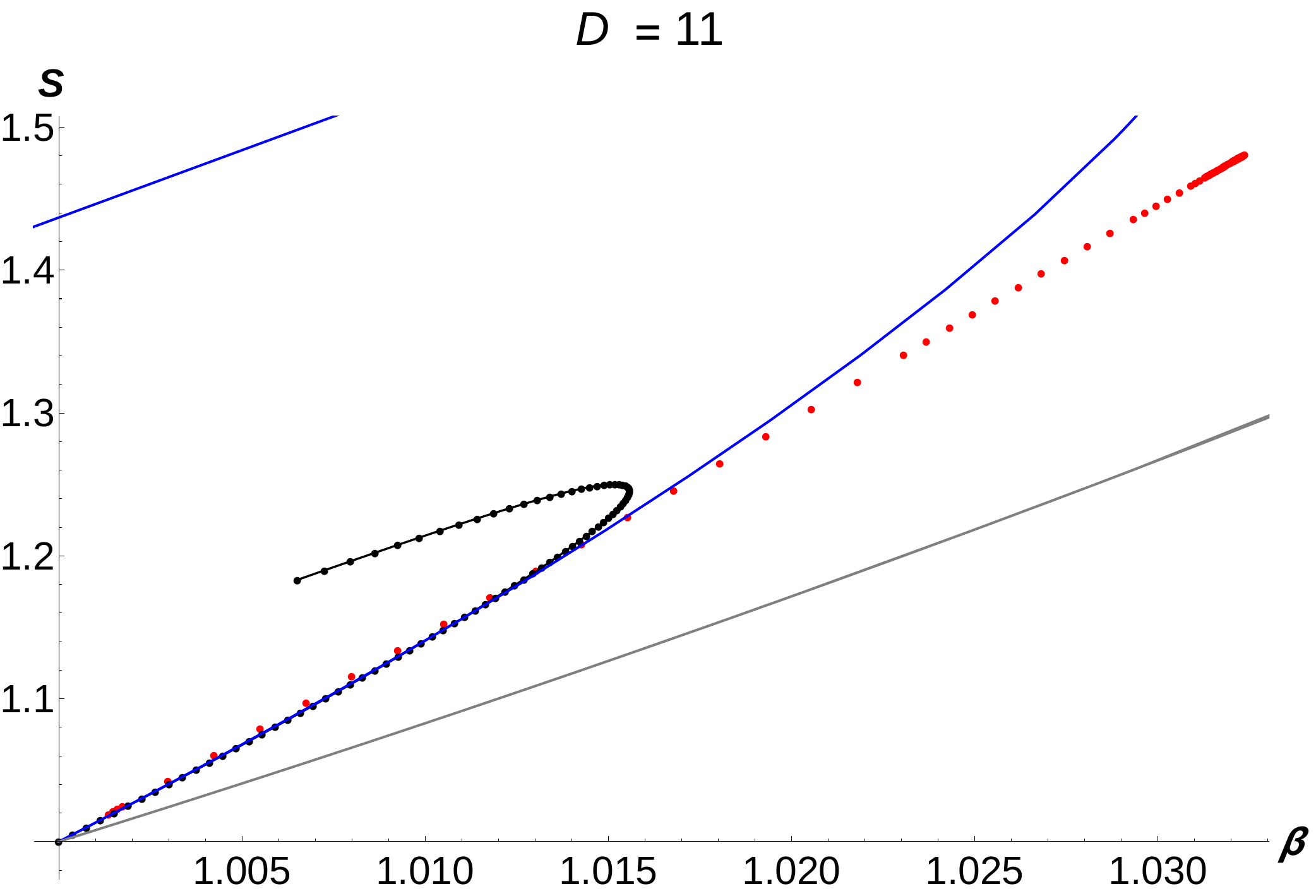}
\includegraphics[width=200pt]{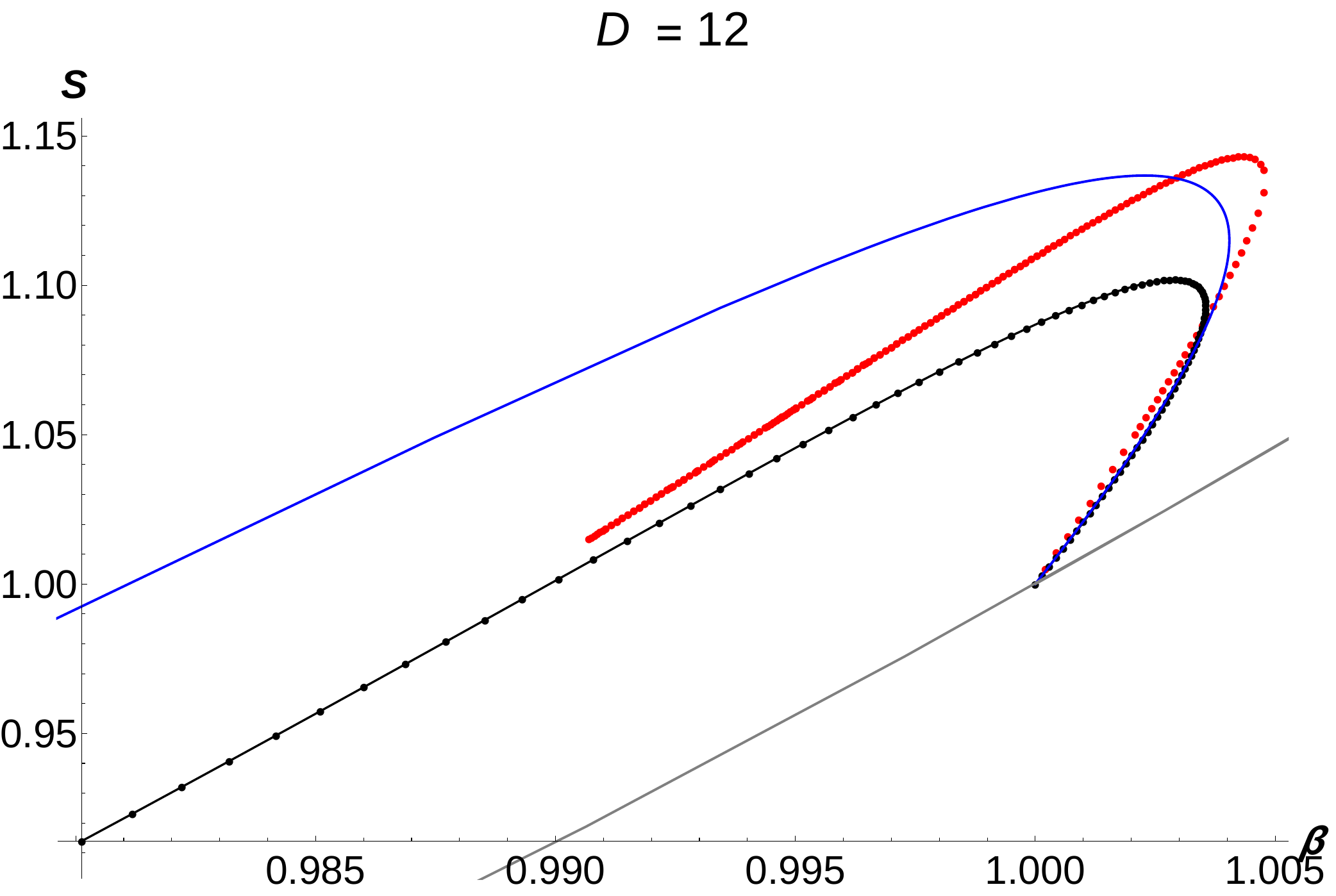}
\hfill\break
\hfill\break
\includegraphics[width=200pt]{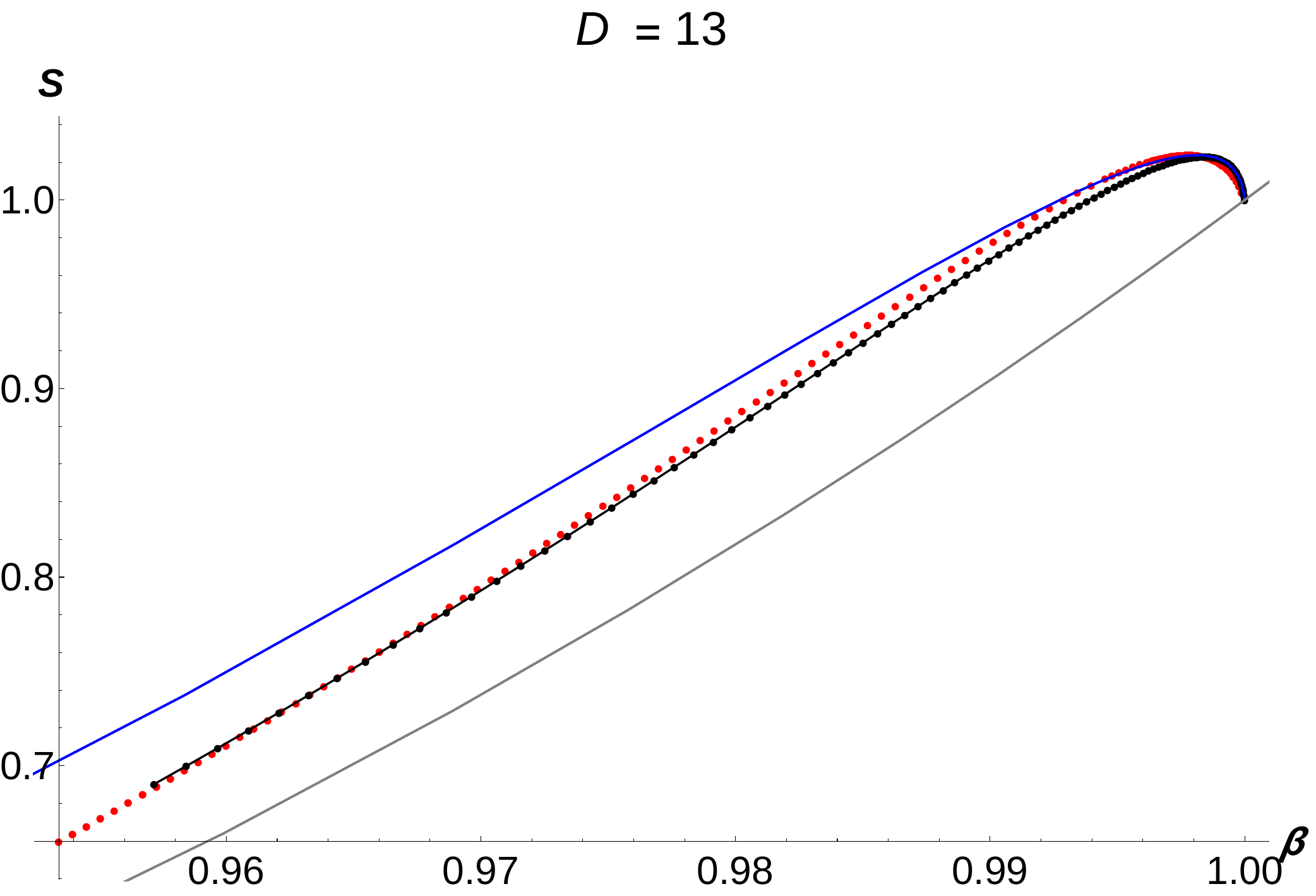}
\includegraphics[width=200pt]{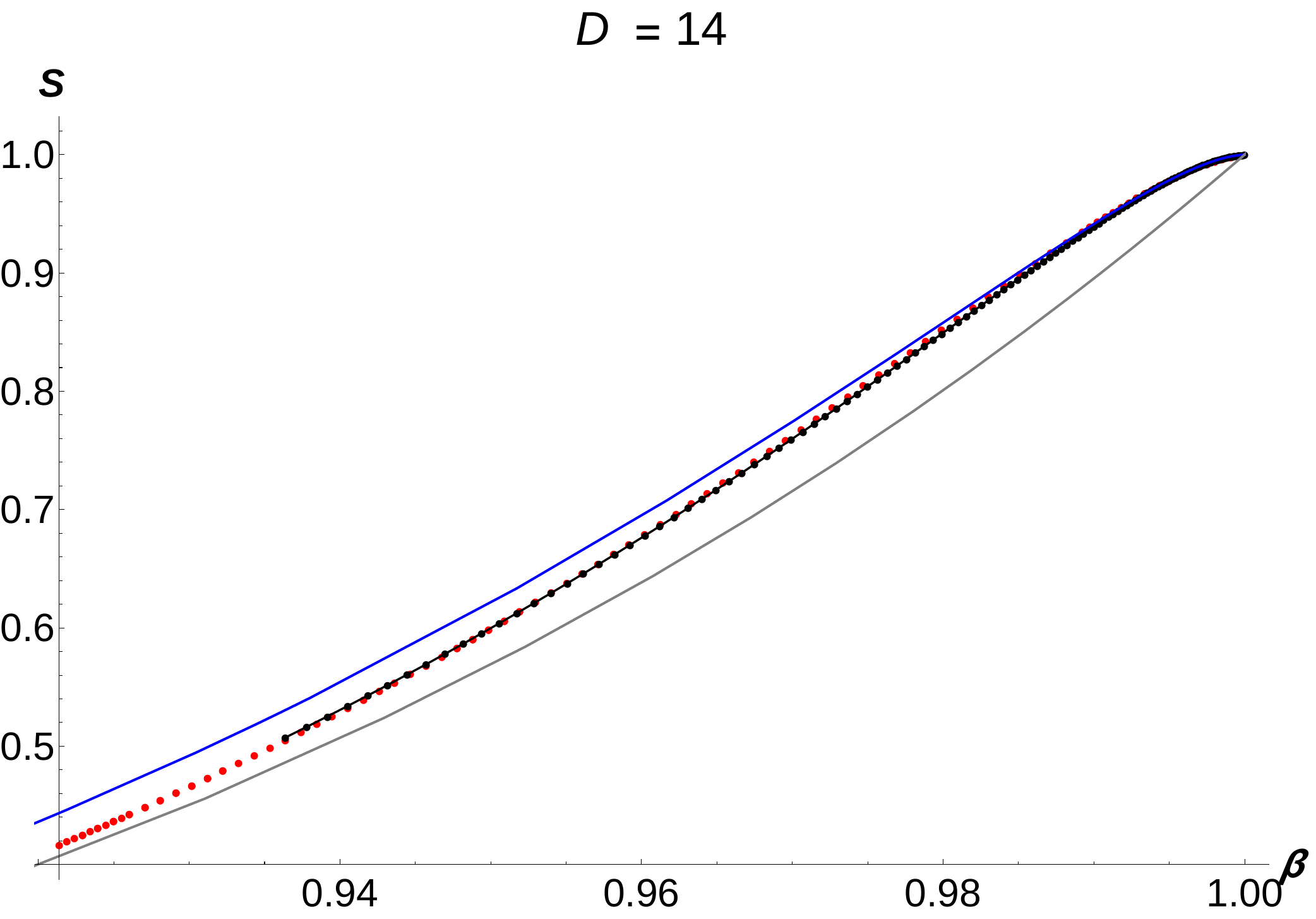}
\caption{\small Entropy $\mathbf{S}$ vs.\ inverse temperature $\bm\beta$. Black dots: large-$D$ numerical results for NUBS.  Blue solid: large-$D$ perturbative solution for NUBS. Red dots: finite-$D$ full-numerical NUBS in \cite{Figueras:2012xj}. Gray solid: uniform black string ($\mathbf{S}_{\textrm{UBS}}= \bm{\beta}^{n+1}$). From the branching point at $\mathbf{S}=\bm{\beta}=1$, we see that $\mathbf{S}$ increases when $D<13.6$ and decreases when $D>13.6$.}\label{fig:BetaA}
\end{center}
\end{figure}

The curves for the relative binding energy $\mathbf{n}$ in fig.~\ref{fig:BetaTension} reproduce the main qualitative features previously found for NUBS. However, they end at $\mathbf{n}=0$ (zero tension) while the actual curves for NUBS branches terminate at positive, non-zero tension, where, at least in $D=5,6$, they merge with black hole phases in a spiralling way \cite{Kleihaus:2006ee,Kalisch:2015via,Kalisch:2016fkm}. These spirals are a feature controlled by the critical self-similar solution at the static merger transition \cite{Kol:2002xz}, which does not seem to be captured by the large-$D$ effective equations for black strings.	It may be visible, though, in large-$D$ studies aimed closer to the self-similar solution.

\begin{figure}[H]
\begin{center}
\includegraphics[width=200pt]{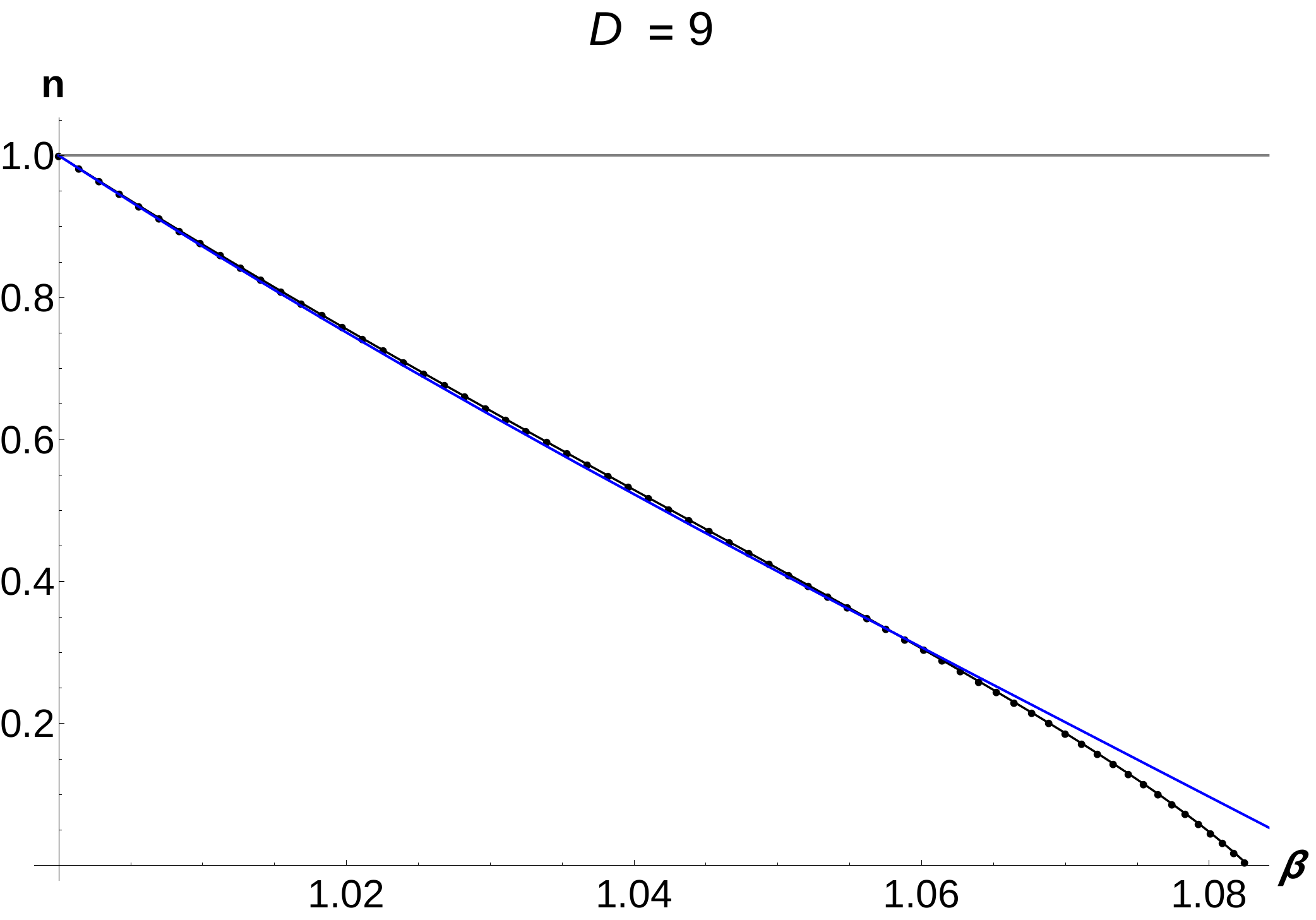}
\includegraphics[width=200pt]{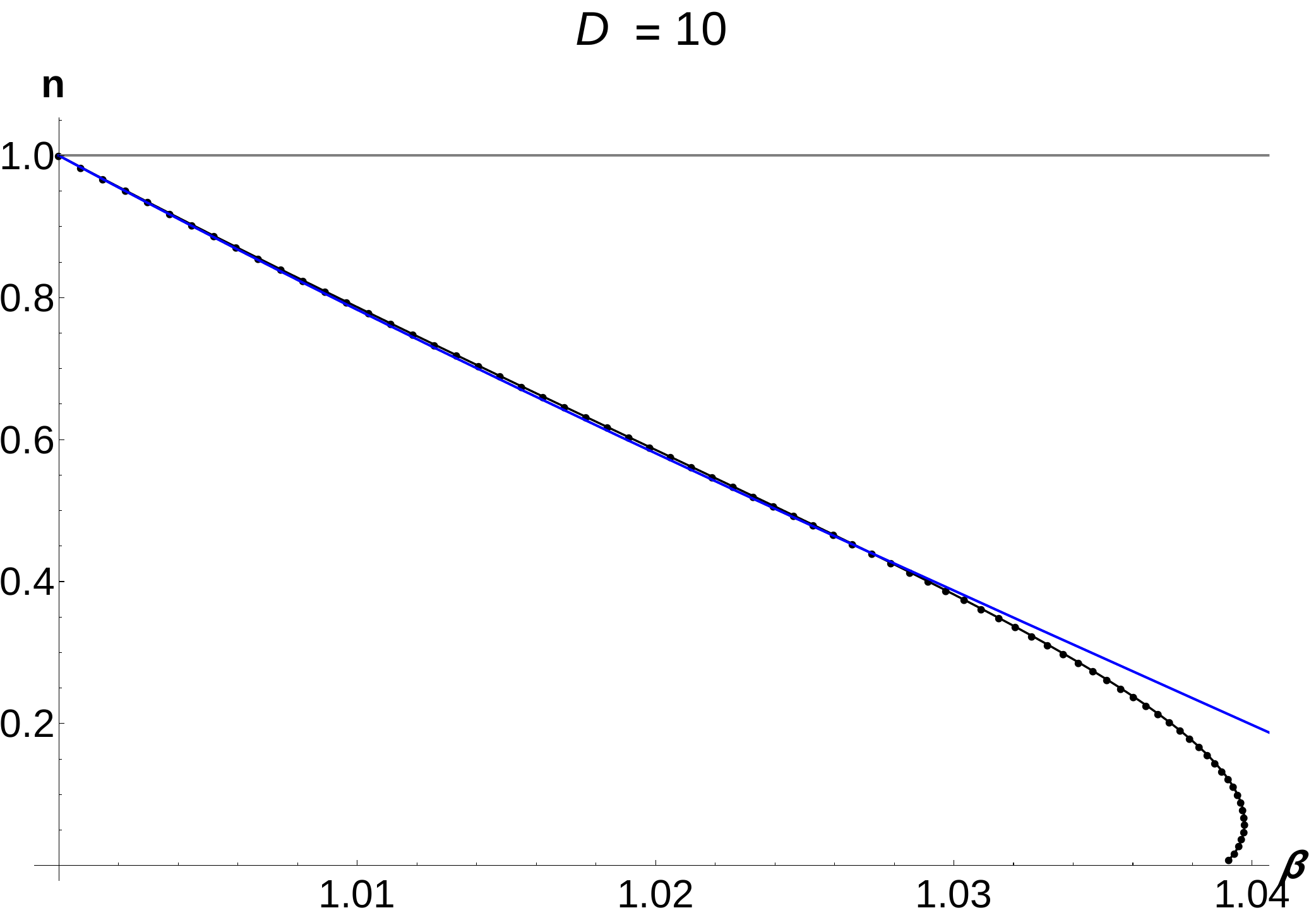}
\hfill\break
\hfill\break
\includegraphics[width=200pt]{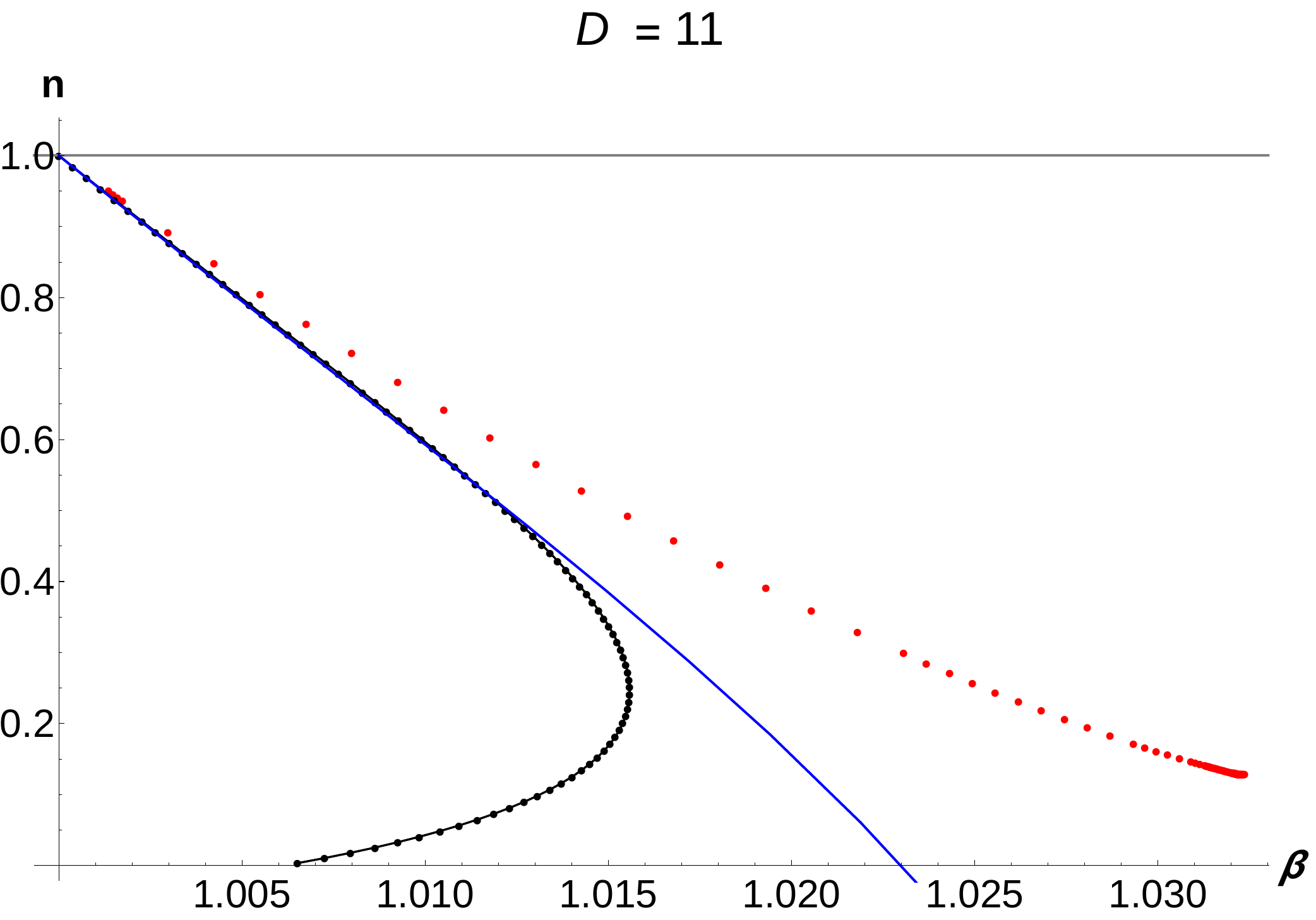}
\includegraphics[width=200pt]{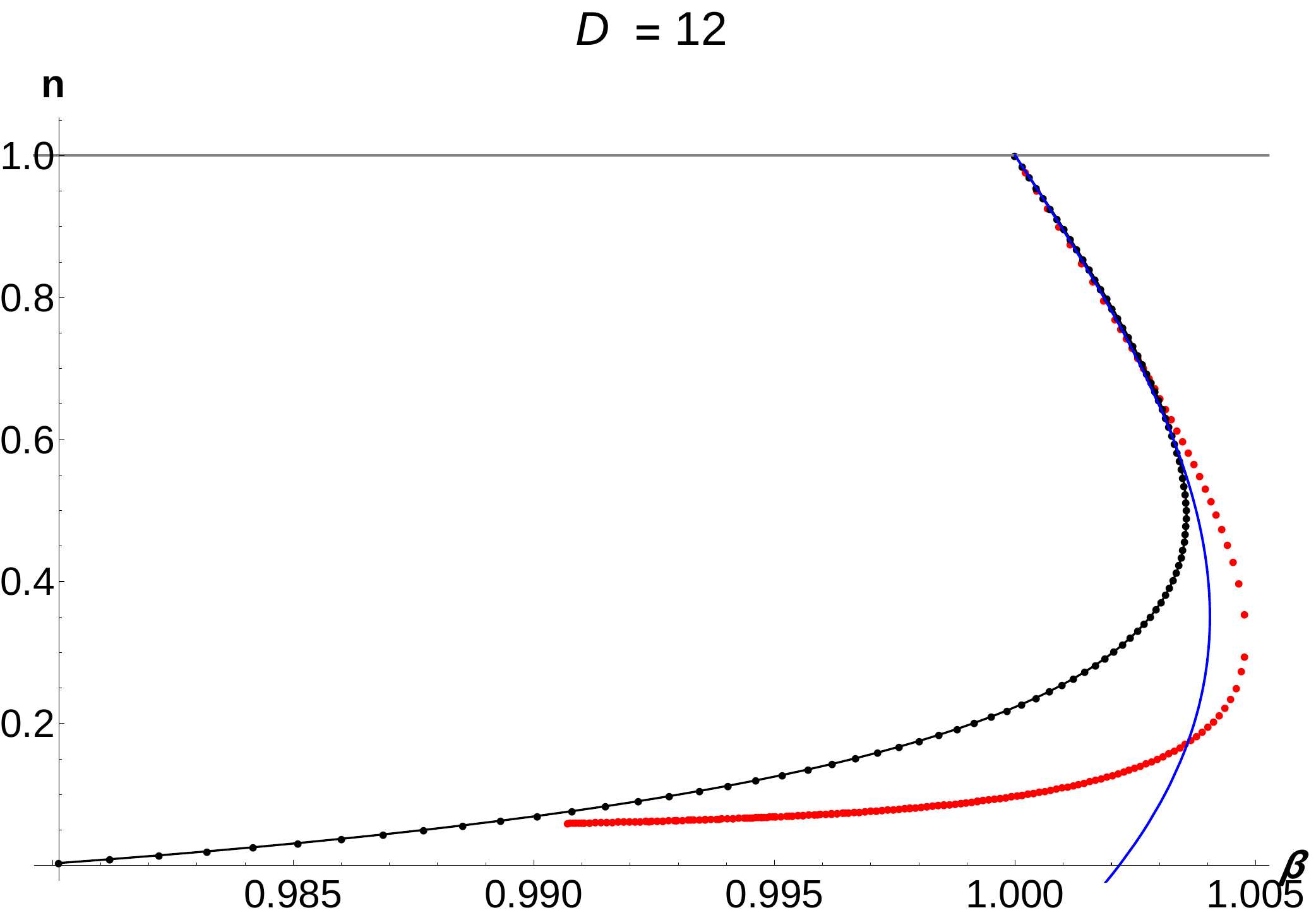}
\hfill\break
\hfill\break
\includegraphics[width=200pt]{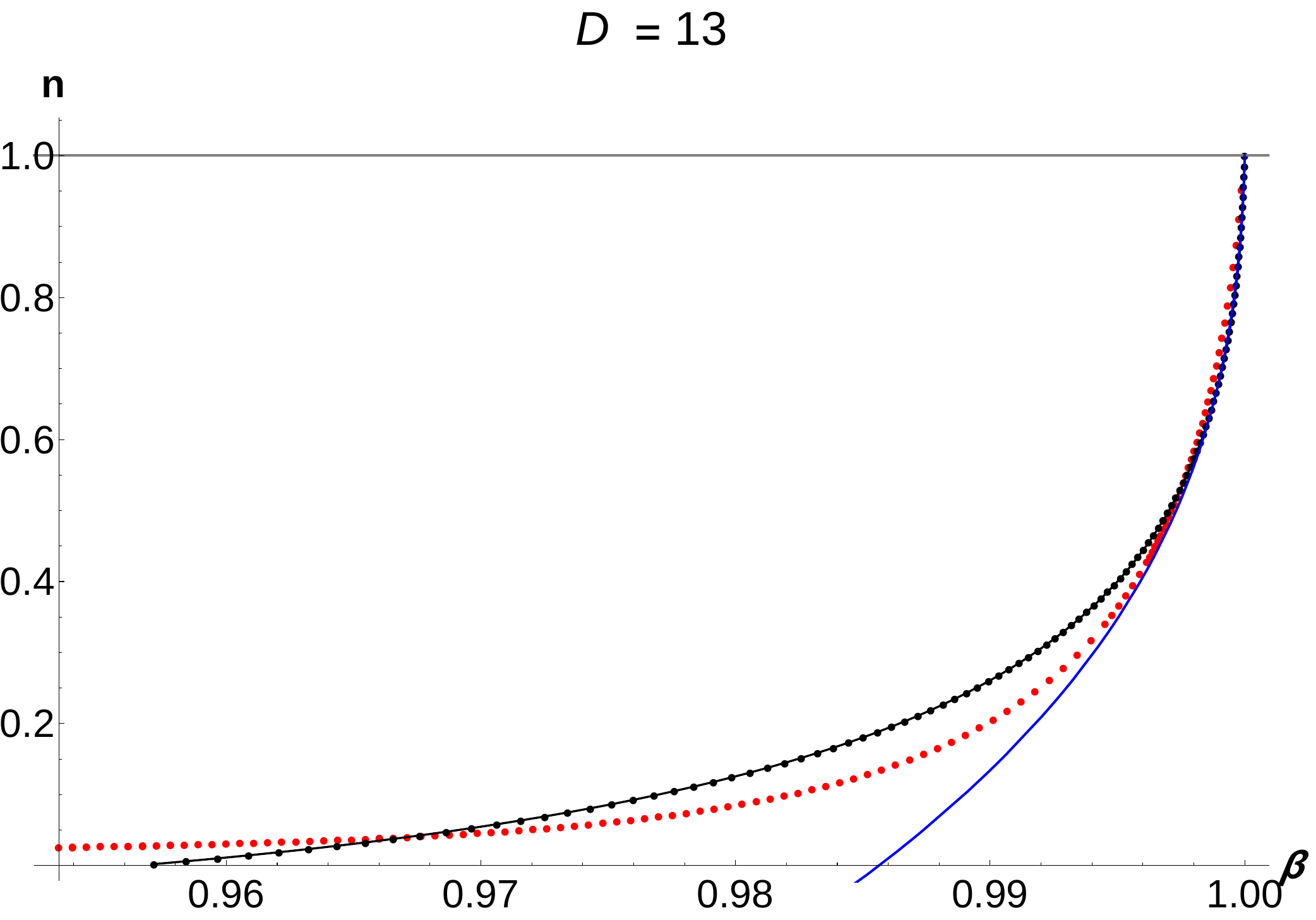}
\includegraphics[width=200pt]{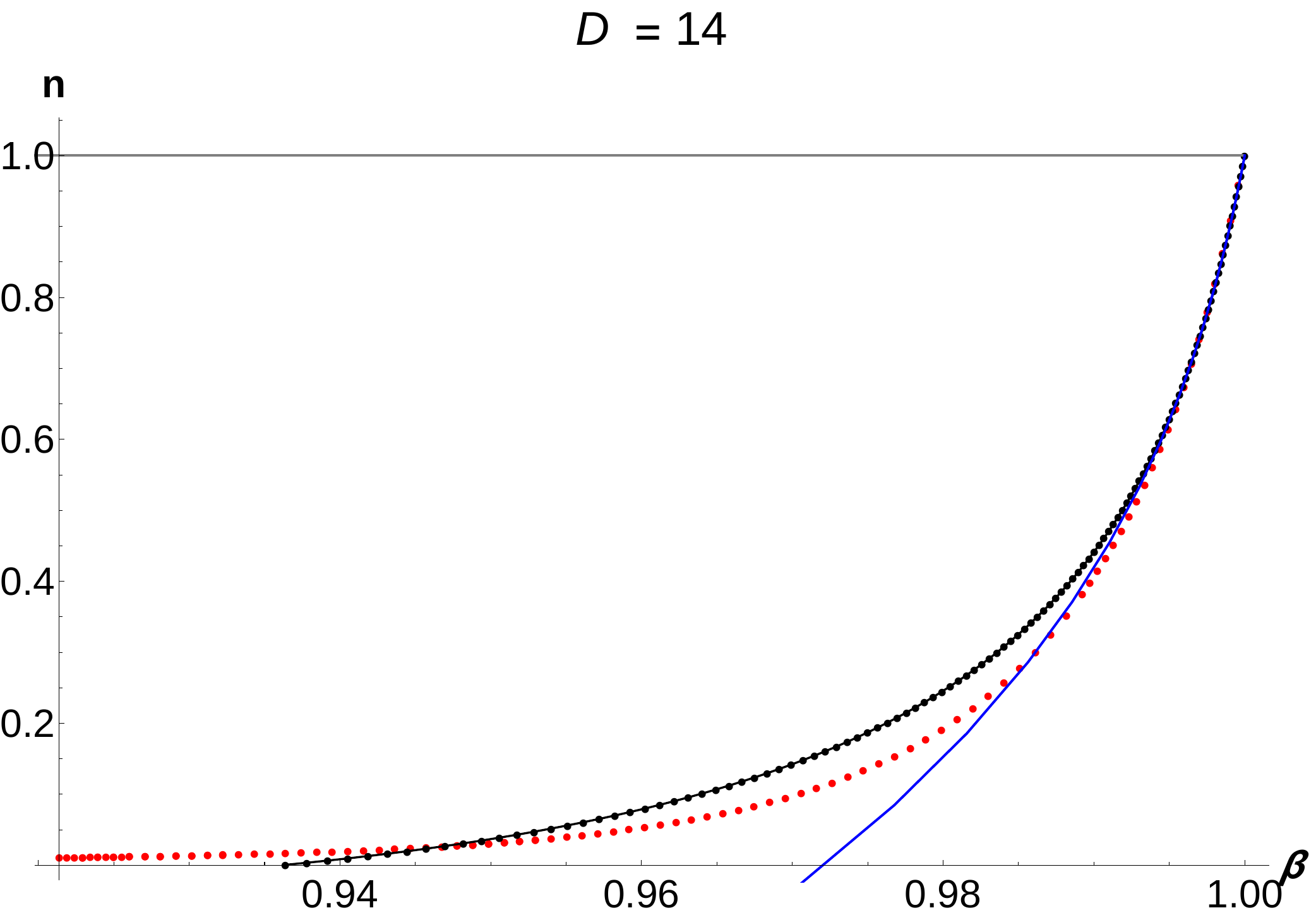}
\caption{\small Relative binding energy $\mathbf{n}$ (tension per unit mass and length) vs.\ inverse temperature $\bm\beta$.  Black dots: large-$D$ numerical results for NUBS.  Blue solid: large-$D$ perturbative solution for NUBS. The line $\mathbf{n}_{\textrm{UBS}}=1$ corresponds to uniform black strings. }\label{fig:BetaTension}
\end{center}
\end{figure}

\begin{figure}[H]
\begin{center}
\includegraphics[width=200pt]{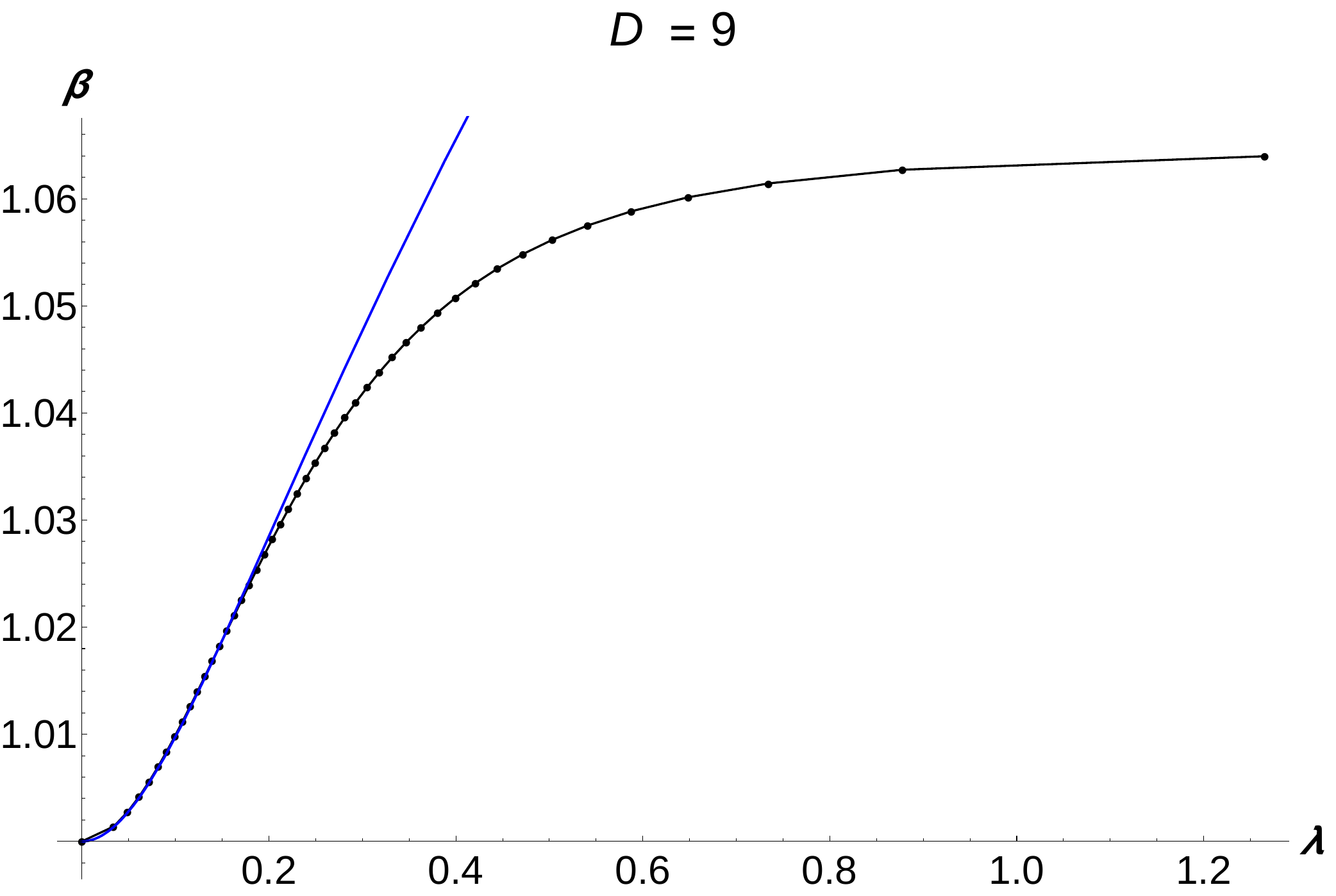}
\includegraphics[width=200pt]{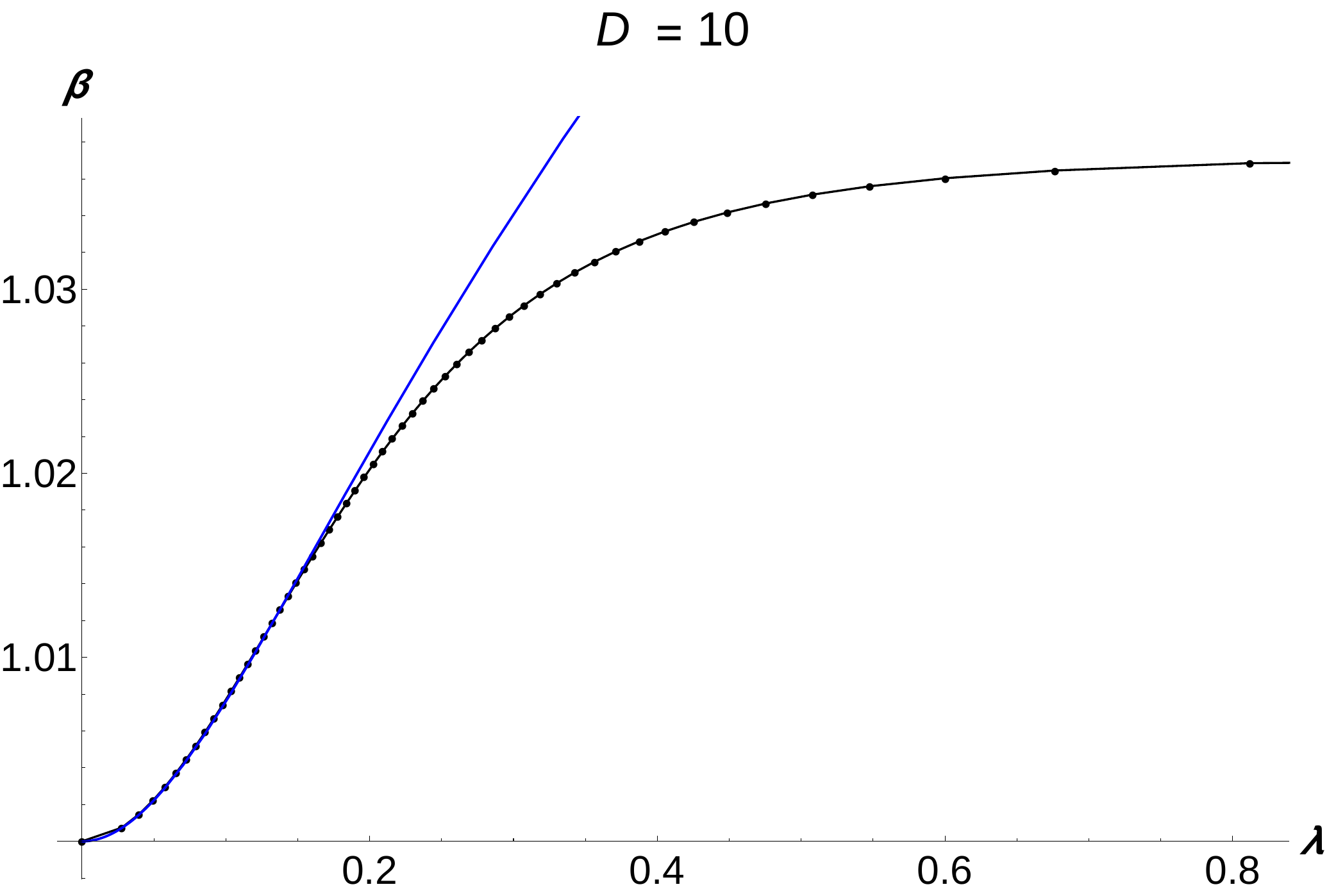}
\hfill\break
\hfill\break
\includegraphics[width=200pt]{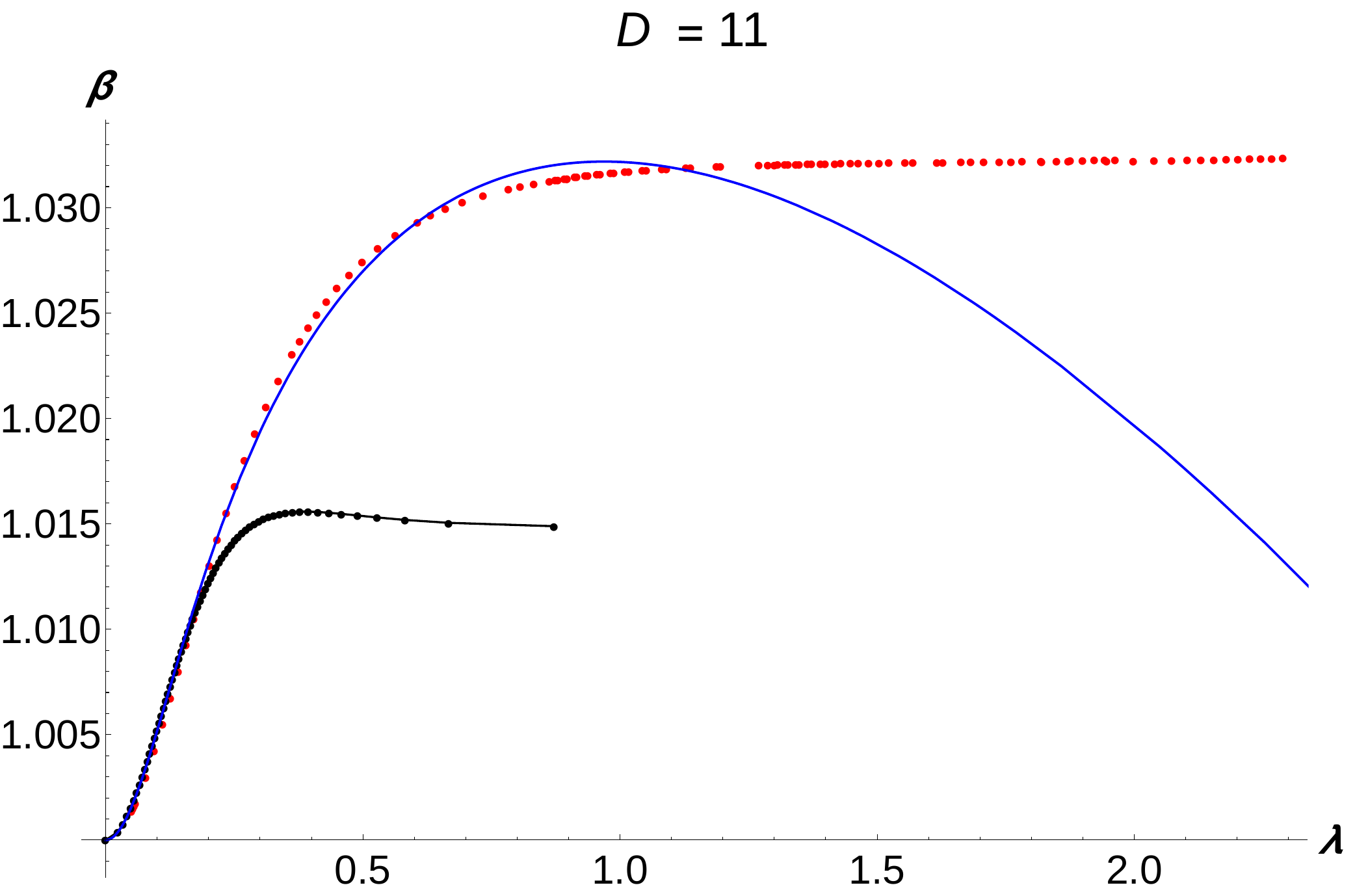}
\includegraphics[width=200pt]{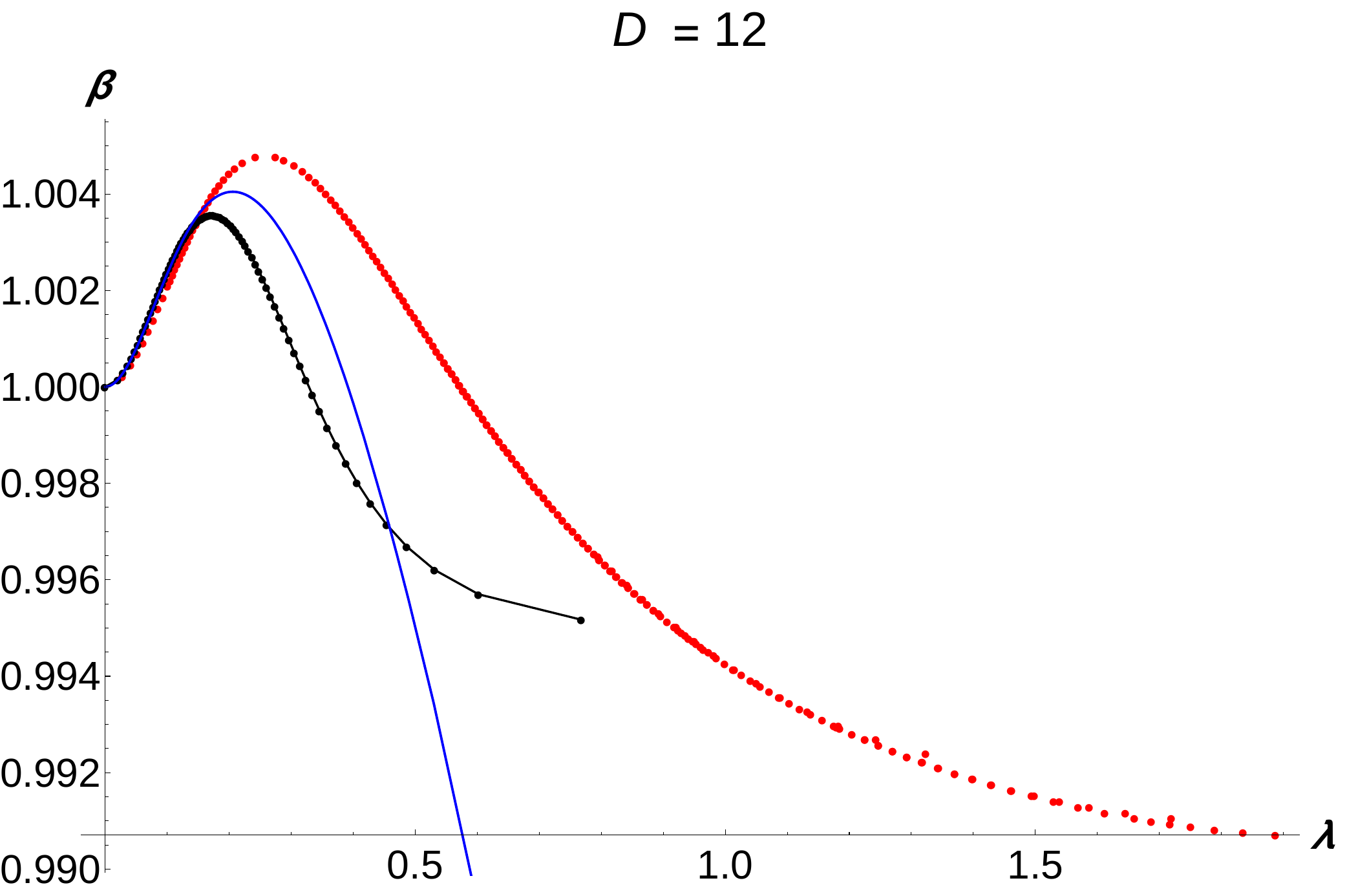}
\hfill\break
\hfill\break
\includegraphics[width=200pt]{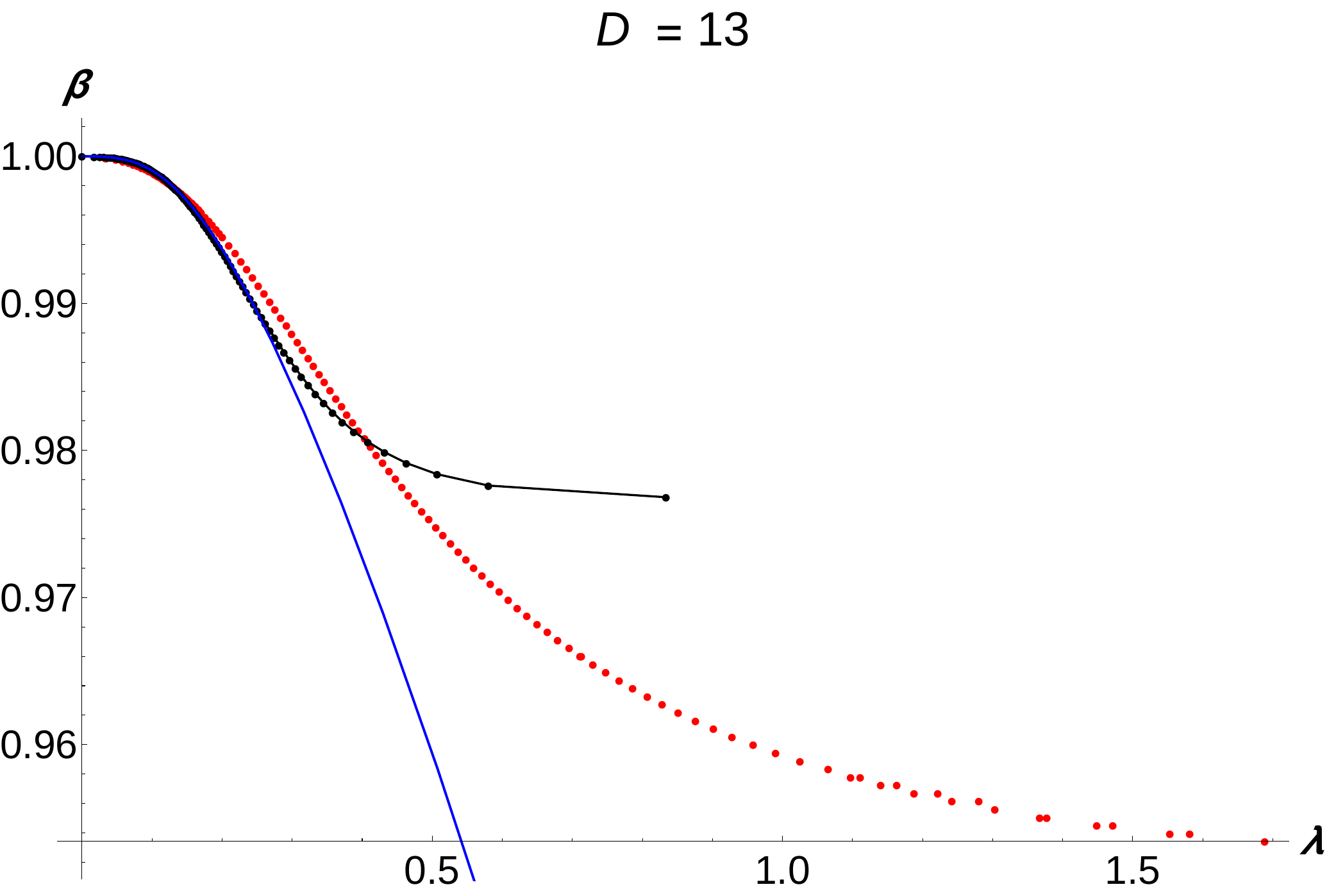}
\includegraphics[width=200pt]{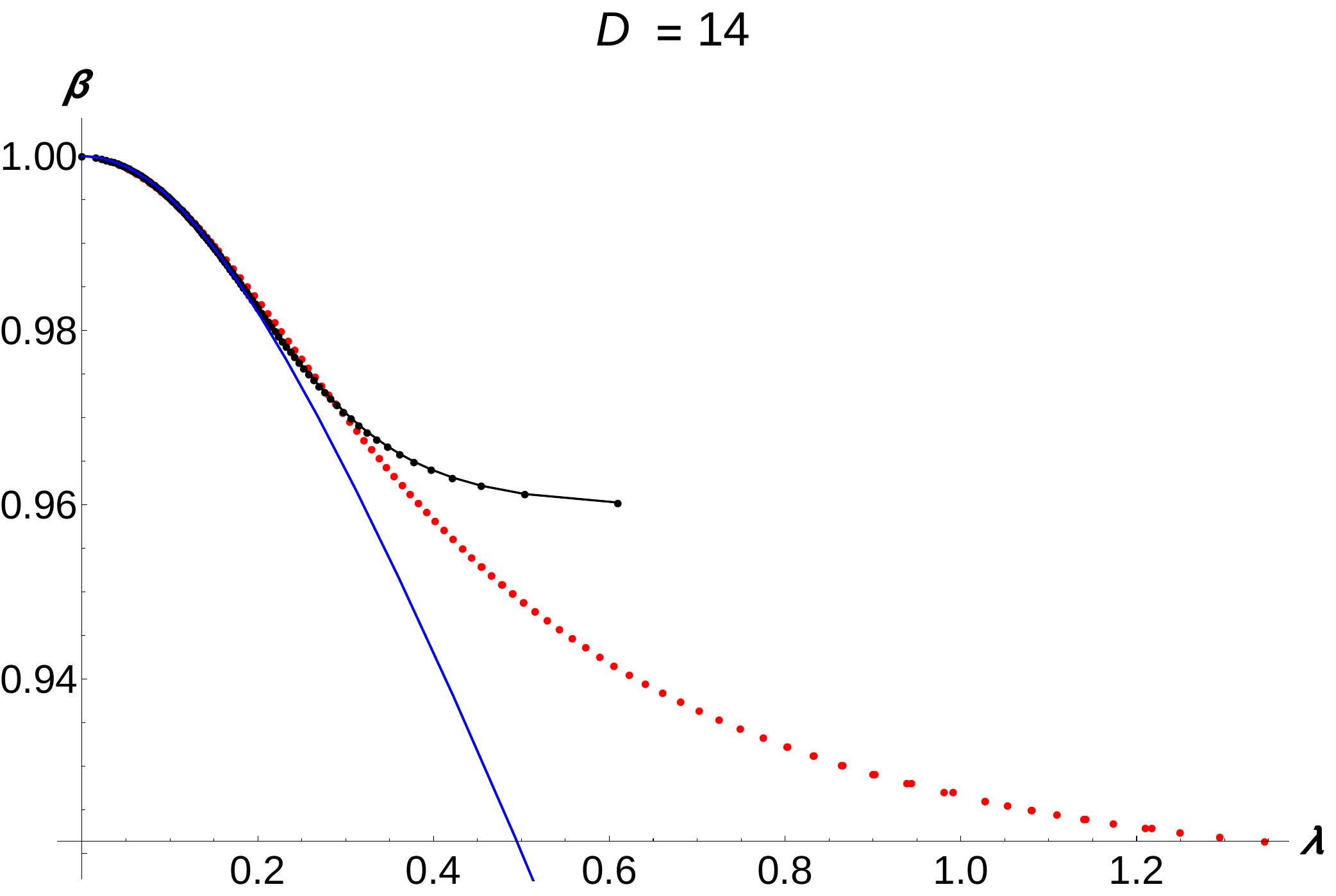}
\caption{\small Inverse temperature $\bm\beta$ vs.\ deformation parameter $\lambda$.  Black dots: large-$D$ numerical results for NUBS.  Blue solid: large-$D$ perturbative solution for NUBS. Red dots: finite-$D$ full-numerical NUBS in \cite{Figueras:2012xj}}\label{fig:LambdaBeta}
\end{center}
\end{figure}

Overall, these diagrams exhibit the main qualitative features of the NUBS branches of static solutions, both above and below the critical dimension $D_*=13.6$. 
The quantitative accuracy is excellent at $D=13$ and higher, but worsens as $D$ gets smaller. Nevertheless, the position of the GL point in all these diagrams is excellently reproduced even down to $D=6$ \cite{Emparan:2015rva}.

\subsection{Dynamics}\label{subsec:stabphases}

Ref.~\cite{Emparan:2015gva} exhibited numerical simulations of the evolution of unstable black strings in the limit $D\to\infty$. They showed that evolution following the LO equations always ends at stable NUBS.
When we incorporate NLO corrections (using the inclusive approach) the outcome of the evolution depends on the thickness of the initial uniform black string.

Black strings with $\mathbf{M}$ not much below the GL instability proceed directly towards stable NUBS with the same value of $\mathbf{M}$. However, when we consider lighter strings, they develop larger inhomogeneity until the evolution breaks down. It would be appealing to attribute this breakdown to a singular pinch-down to zero size of the black string horizon, which would be naturally followed by a (non-classical) transition to a BH configuration. However, we cannot distinguish this effect from a more banal breakdown of the large-$D$ expansion as the inhomogeneity grows too large. Indeed, we expect that this is the reason that the evolutions break down when the mass of the black string is in a range  $\mathbf{M}>\mathbf{M}_{\textrm{min}}(D)$, where there exist stable NUBS that are natural endpoints. 

Observe that the breakdown can happen only in the inclusive approach; the sequential method cannot lead to any such breakdown, only to the appearance of unphysical solutions that would have negative tension, such as we have constructed above. That is, time evolution in the sequential approach can be regarded as a particular relaxation approach to obtain static solutions.

We conclude that the numerical time evolution of black strings in the large-$D$ expansion can be useful to verify the stability of NUBS when these are not too light. However, the (inclusive) large-$D$ approach breaks down when the mass $\mathbf{M}$ of the black strings becomes small enough, even within the range of existence of stable NUBS.

\section{Final remarks and outlook}\label{sec:fin}

It is remarkable that, despite a qualitative change at a finite, critical dimension---often an impassable barrier to a perturbative series---, the large-$D$ expansion is able to correctly capture physics at and even below this dimension. In particular, we have been able to compute $D_*$ analytically as it is manifested in several different magnitudes, with all the 3NLO results \eqref{massDstar}, \eqref{areaDstar} and \eqref{dynDstar} converging on the value
\beq
D_*=13.6\,.
\eeq

By solving the effective equations to increasingly high perturbative order, we have performed a thorough study of the properties---static, thermodynamic, and dynamic---of non-uniform black strings in a Kaluza-Klein circle in a very wide range of dimensions. Even though we have often resorted to numerical integration of the equations, there are still useful benefits to this approach, as compared to direct full numerical integration of Einstein's equations. One advantage is that the effective equations, even if lengthy at high perturbative orders, are uncomplicated from a conceptual viewpoint. There are no constraints to be solved, and issues such as gauge invariance and gauge fixing are absent---they are dealt with once and for all when the effective equations are derived. A second, even more substantial advantage, lies in the possibility of using a sequential construction (sec.~\ref{sec:nums}) in which, once the corrections to a given order are computed, then one can combine them to directly generate solutions where $D$ is a continuous parameter that can be freely varied. 

A natural question is whether some of the rich structure observed in the time evolutions in $D=5$ in \cite{Lehner:2010pn}, in particular the cascading behavior at late times, may be reproduced by evolving the effective large-$D$ equations, which are computationally relatively simple to solve. After all, as we have seen, when sufficiently high-order corrections are included these equations appear to correctly capture many qualitative properties of NUBS in $D<D_*$. Unfortunately, we have found that the time evolution breaks down too early in the development of inhomogeneity to see this. Our simulations do not exhibit any sign of these cascading structures, nor of the self-similar shrinking of the thin tubes that connect the larger blobs on the string in \cite{Lehner:2010pn}. Indeed, this behavior seems to depend crucially on scaling and homogeneity properties of the equations that are washed away when taking $D\to\infty$. So all those detailed features may be inaccessible to the large-$D$ expansion.

Lacking so far in our investigation of black objects in a Kaluza-Klein circle are the phases of localized black holes. The reason is that, like with numerical analyses, the presence of `exposed' sections of the symmetry axis that are not covered by a horizon requires an approach different than for black string phases. Nevertheless, we expect that the large-$D$ expansion can be a useful means for their investigation. It is straightforward to construct, via a large-$D$ matched asymptotic expansion, solutions for localized black holes of increasing size up until they almost fill up the axis of the compact circle, except for a fraction $\sim 1/D$ of it.\footnote{This is a large-$D$ version of the constructions in \cite{Harmark:2003yz,Gorbonos:2004uc}.} Black holes that fill up a larger fraction of the axis and then reach up to the merger configurations,\footnote{These have been numerically studied at finite, low $D$, in \cite{Kudoh:2004hs,Headrick:2009pv,Kalisch:2017bin}} where conifold structures of the type studied in \cite{Kol:2002xz,Kol:2003ja,Emparan:2011ve} appear, also seem to be accessible in a large-$D$ expansion. Putting together all these pieces of information, it might then be possible to obtain a complete characterization of the phases of black objects in a Kaluza-Klein circle across the entire range of dimensions.

We should not finish without discussing the relevance of what we have achieved. One may question the worth of figuring out the properties of black strings in dimensions well above those contemplated in, say, M-theory. This objection is not entirely without force, but we place the lessons of this article elsewhere. We believe that we have not only gained a better understanding of the rich casuistics of the phenomenon of spontaneous symmetry breakdown in a gravitational system---which may be exported to other related systems. We have also shown that it is amenable to very significant simplification, from the fiendish complexity of Einstein's equations, down to the highly tractable large-$D$ effective equations, which---as we have demonstrated---can bring to light many of the most interesting features of the system. Hopefully, these techniques, and what has been learned by developing them, will also be useful in the investigation of other problems.

\section*{Acknowledgments}
We are grateful to Moshe Rozali for useful discussions, and to Jorge Santos for comments that prompted us to revise the first version of this article. We also thank Pau Figueras, Keiju Murata, and Harvey Reall for providing the data used in figs.~\ref{fig:BetaM}, \ref{fig:BetaA}, \ref{fig:BetaTension} and \ref{fig:LambdaBeta}. Much of this work was performed during the Benasque workshops on ``Gravity: New perspectives from strings and higher dimensions'' in July 2015 and July 2017.
This work is partly supported by MEC grants FPA2013-46570-C2-2-P, FPA2016-76005-C2-2-P and FPU15/01414; AGAUR grant 2009-SGR-168; ERC Advanced Grant GravBHs-692951; CPAN CSD2007-00042 Consolider-Ingenio 2010; by the JSPS Program for Advancing Strategic International Networks to Accelerate the Circulation of Talented Researchers ``Mathematical Science of Symmetry, Topology and Moduli, Evolution of International Research Network based on OCAMI''. KT was supported by JSPS Grant-in-Aid for Scientific Research No.26-3387 and MM was supported by FWO grant G092617N.


\addcontentsline{toc}{section}{Appendices}
\appendix

\section{Perturbative static solution}\label{app:pertsol}

\subsection{Wavenumber $k(\epsilon)$}

\begin{eqnarray}\label{kGL4NLO}
k_{\rm GL}=1-\frac{1}{2 n}+\frac{7}{8 n^2}+\frac{-\frac{25}{16}+2 \zeta(3) }{n^3}+\frac{\frac{363}{128}-5
   \zeta(3) }{n^4}\,.
\end{eqnarray}

\begin{eqnarray} \label{kkgl}
\begin{split}
&\frac{k}{k_{\rm GL}} = 1+\epsilon ^2 \left(-\frac{1}{24}+\frac{1}{3 n}+\frac{7}{12 n^2}+\frac{\frac{1}{4}+\frac{5 \zeta(3) }{6}}{n^3}+\frac{\frac{7}{6}+\frac{\pi ^4}{9}+\frac{5   \zeta(3) }{3}}{n^4}\right)\\
   &+\epsilon ^4   \left(-\frac{77}{6912}+\frac{25}{864 n}+\frac{191}{576 n^2}+\frac{2201-118 \zeta(3) }{1728 n^3}+\frac{74595+868 \pi ^4-23580 \zeta(3) }{25920   n^4}\right)\\
   &+\epsilon ^6 \left(-\frac{11063}{3317760}-\frac{991}{138240 n}+\frac{123323}{1658880 n^2}+\frac{1383971-184158 \zeta(3) }{1658880 n^3}
\right.\\
   &\left.\qquad
+\frac{7802075+14147 \pi   ^4-2093535 \zeta(3) }{2073600 n^4}\right)\,.
      \end{split}
\end{eqnarray}

\subsection{$m(z)$ expansion coefficients}

We only give the results up to NNLO and $\epsilon^6$ (recall $\mu_i$ and $\nu_i$ are multiplied by $\epsilon^i$). Higher orders can be found in the ancillary \textsl{Mathematica} files.

We have set
\begin{dmath} 
\mu_{1} = 1
\end{dmath}
as a definition of the perturbative expansion parameter $\epsilon$.

\begin{dmath} 
\mu_{2} = \frac{1}{6}+\frac{1}{3 n}-\frac{1}{2 n^2}+\left(\frac{19}{864}-\frac{5}{48 n}+\frac{53}{96 n^2}\right) \epsilon ^2+\left(\frac{731}{138240}-\frac{907}{41472 n}+\frac{6863}{138240 n^2}\right) \epsilon ^4+ \ord{\epsilon^{6}}
\end{dmath}

\begin{dmath} 
\mu_{3} = \frac{1}{96}+\frac{1}{6 n}-\frac{11}{48 n^2}+\left(\frac{163}{46080}+\frac{23}{2880 n}-\frac{347}{2880 n^2}\right) \epsilon ^2+ \ord{\epsilon^{4}}
\end{dmath}

\begin{dmath} 
\mu_{4} = \frac{1}{4320}+\frac{1}{80 n}+\frac{79}{480 n^2}+\left(\frac{233}{1555200}+\frac{4099}{777600 n}-\frac{16099}{1555200 n^2}\right) \epsilon ^2+ \ord{\epsilon^{4}}
\end{dmath}

\begin{dmath} 
\mu_{5} = \frac{1}{414720}+\frac{11}{25920 n}+ \ord{\epsilon^{2}}
\end{dmath}

\begin{dmath} 
\mu_{6} = -\frac{1}{29030400}-\frac{353}{14515200 n}+\frac{105223}{29030400 n^2}+ \ord{\epsilon^{2}}
\end{dmath}

\subsection{$p(z)$ expansion coefficients}

\begin{dmath} 
\nu_{1} = -1+\frac{1}{2 n}+\frac{1}{8 n^2}+\left(\frac{1}{24}-\frac{19}{48 n}-\frac{29}{192 n^2}\right) \epsilon^2+\left(\frac{77}{6912}-\frac{595}{13824 n}-\frac{19409}{55296 n^2}\right) \epsilon^4+ \ord{\epsilon^{6}}
\end{dmath}

\begin{dmath} 
\nu_{2} = -\frac{1}{3}+\frac{1}{8 n^2}+\left(-\frac{13}{432}+\frac{7}{54 n}-\frac{3785}{3456 n^2}\right) \epsilon^2+\left(-\frac{1043}{207360}+\frac{247}{25920 n}-\frac{3917}{552960 n^2}\right) \epsilon^4+ \ord{\epsilon^{6}}
\end{dmath}

\begin{dmath} 
\nu_{3} = -\frac{1}{32}-\frac{23}{64 n}+\frac{289}{256 n^2}+\left(-\frac{143}{15360}+\frac{511}{30720 n}-\frac{543}{8192 n^2}\right) \epsilon ^2+ \ord{\epsilon^{4}}
\end{dmath}

\begin{dmath} 
\nu_{4} = -\frac{1}{1080}-\frac{23}{540 n}-\frac{3941}{8640 n^2}+\left(-\frac{109}{194400}-\frac{1501}{97200 n}+\frac{48757}{518400 n^2}\right) \epsilon ^2+ \ord{\epsilon^{4}}
\end{dmath}

\begin{dmath} 
\nu_{5} = -\frac{1}{82944}-\frac{109}{55296 n}+\frac{4379}{221184 n^2}+ \ord{\epsilon^{2}}
\end{dmath}

\begin{dmath} 
\nu_{6} = \frac{1}{4838400}+\frac{43}{302400 n}-\frac{933859}{38707200 n^2}+ \ord{\epsilon^{2}}
\end{dmath}

\section{Thermodynamics}\label{app:thermo}

The equations we use for the thermodynamical variables as a function of $m(z)$ are

\begin{dmath} 
M(z) = m_0(z)+\frac{1}{n} \left(m_1(z)-2 m_0''(z)\right)+\ord{\frac{1}{n^2}}
\end{dmath}

\begin{dmath} 
A_h(z) = m_0(z)+\frac{1}{n} \left(\frac{m_0'(z){}^2}{2 m_0(z)}+m_0''(z) \log \left(m_0(z)\right)+m_1(z)+m_0(z) \log \left(m_0(z)\right)\right)+\ord{\frac{1}{n^2}}
\end{dmath}

\begin{dmath} 
\tau(z) = \left(-\frac{m_0'(z){}^2}{m_0(z)}+m_0''(z)+m_0(z)\right)+\frac{1}{n} \left(-\frac{3 m_0'(z){}^2 m_0''(z)}{m_0(z){}^2}-\frac{4 m_0'(z){}^2 m_0''(z) \log \left(m_0(z)\right)}{m_0(z){}^2}+\frac{m_0'(z){}^4}{m_0(z){}^3}+\frac{m_1(z) m_0'(z){}^2}{m_0(z){}^2}-\frac{3 m_0'(z){}^2}{m_0(z)}-\frac{2 m_1'(z) m_0'(z)}{m_0(z)}+\frac{2 m_0'(z){}^4 \log \left(m_0(z)\right)}{m_0(z){}^3}-\frac{2 m_0'(z){}^2 \log \left(m_0(z)\right)}{m_0(z)}+\frac{2 m_0''(z){}^2}{m_0(z)}+m_0''(z)+m_1''(z)+\frac{2 m_0''(z){}^2 \log \left(m_0(z)\right)}{m_0(z)}+2 m_0''(z) \log \left(m_0(z)\right)+m_1(z)\right)+\ord{\frac{1}{n^2}}
\end{dmath}

\begin{dmath} 
\hat \kappa(z) = 1+\frac{1}{n} \left(\frac{m_0'(z){}^2}{2 m_0(z){}^2}-\frac{m_0''(z)}{m_0(z)}-\log \left(m_0(z)\right)\right)+\ord{\frac{1}{n^2}}
\end{dmath}

\begin{dmath} 
\sR_h(z) = m_0(z)+\frac{1}{n} \left(m_0''(z) \log \left(m_0(z)\right)+m_1(z)\right)+\ord{\frac{1}{n^2}}
\end{dmath}

\section{Thermodynamical quantities and quasinormal mode}

In this appendix we present, for the perturbative solutions obtained in appendix~\ref{app:pertsol}, the values for 
\beq
\mathcal{M}=\frac1{L_{GL}}\int_{-L/2}^{L/2} dz\, M(z)\,,\qquad \mathcal{S}=\frac1{L_{GL}}\int_{-L/2}^{L/2} dz\, S(z)\,.
\eeq
These are normalized so that for the UBS at the GL threshold we have $\mathcal{M}=1$, $\mathcal{S}=1$ at all $n$. 

With these quantities, we obtain the mass and entropy \eqref{calma} as
\beq\label{mal}
\mathbf{M}=\lp\frac{L_{GL}}{L}\rp^{n+1}\mathcal{M}\,,\qquad 
\mathbf{S}=\lp\frac{L_{GL}}{L}\rp^{n+2}\mathcal{S}\,.
\eeq	
One reason to separately have expressions for $\mathcal{M}$ and $\mathcal{A}$ is that they are not dominated at large $n$ by the factors $\sim (L_{GL}/L)^n$.

Here we show our results up to NNLO and $\epsilon^{6}$. Calculations up to 4NLO and $\epsilon^8$ can be found in the ancillary \textsl{Mathematica} files.

\begin{dmath} 
\mathcal{M} = 1+\left(\frac{1}{24}-\frac{1}{3 n}-\frac{13}{12 n^2}\right) \epsilon ^2+\left(\frac{89}{6912}-\frac{49}{864 n}-\frac{175}{576 n^2}\right) \epsilon ^4+\left(\frac{14383}{3317760}-\frac{203}{46080 n}-\frac{147013}{1658880 n^2}\right) \epsilon ^6+ \ord{\epsilon^{8}} 
\end{dmath}

\begin{dmath} 
\mathcal{S} = 1+\left(\frac{1}{24}-\frac{1}{3 n}-\frac{13}{12 n^2}\right) \epsilon ^2+\left(\frac{89}{6912}-\frac{5}{108 n}-\frac{217}{576 n^2}\right) \epsilon ^4+\left(\frac{14383}{3317760}+\frac{31}{138240 n}-\frac{172933}{1658880 n^2}\right) \epsilon ^6+ \ord{\epsilon^{8}} 
\end{dmath}

\begin{dmath} 
\tau = 1+\left(-\frac{1}{2}-\frac{1}{n}-\frac{1}{2 n^2}\right) \epsilon ^2+\left(-\frac{1}{72}-\frac{5}{24 n}-\frac{11}{9 n^2}\right) \epsilon ^4+\left(\frac{535}{165888}-\frac{1513}{82944 n}-\frac{78589}{165888 n^2}\right) \epsilon ^6+ \ord{\epsilon^{8}} 
\end{dmath}

\begin{dmath} 
\hat \kappa = 1+\frac{\epsilon ^2}{2 n^2}+\left(-\frac{1}{96 n}+\frac{25}{288 n^2}\right) \epsilon ^4+\left(-\frac{29}{6912 n}+\frac{977}{165888 n^2}\right) \epsilon ^6+ \ord{\epsilon^{8}} 
\end{dmath}

\begin{dmath} 
\lambda = \left(\frac{1}{n}+\frac{\pi ^2}{6 n^3}\right) \epsilon +\frac{\epsilon ^2}{n^2}+\left(\frac{17}{96 n}+\frac{1}{2 n^2}+\frac{\frac{85}{48}+\frac{9 \pi ^2}{64}}{n^3}\right) \epsilon ^3+\left(\frac{17}{48 n^2}+\frac{1}{n^3}\right) \epsilon ^4+\left(\frac{5299}{103680 n}+\frac{4051}{12960 n^2}+\frac{\frac{4817}{2880}+\frac{37411 \pi ^2}{622080}}{n^3}\right) \epsilon ^5+\left(\frac{55397}{414720 n^2}+\frac{10397}{12960 n^3}\right) \epsilon ^6+ \ord{\epsilon^{8}} 
\end{dmath}

Then
\begin{dmath}
\mathbf{M}=1+n \epsilon ^2\left(-\frac{1}{24}+\frac{1}{3 n}+\frac{7}{12
   n^2}\right) +n^2 \epsilon ^4\left(\frac{1}{1152}-\frac{179}{6912 n}+\frac{2}{27 n^2}\right) +n^3 \epsilon ^6\left(-\frac{1}{82944}+\frac{131}{165888 n}-\frac{37883}{3317760
   n^2}\right) + \ord{\epsilon^{8}}
\end{dmath}

\begin{dmath}
\mathbf{S}=1+n \epsilon ^2\left(-\frac{1}{24}+\frac{7}{24
   n}+\frac{11}{12 n^2}\right) +n^2 \epsilon ^4\left(\frac{1}{1152}-\frac{167}{6912 n}+\frac{9}{256 n^2}\right) +n^3 \epsilon ^6\left(-\frac{1}{82944}+\frac{125}{165888 n}-\frac{10601}{1105920
   n^2}\right) + \ord{\epsilon^{8}}
\end{dmath}

\begin{dmath}
\bm{\beta}=1+\epsilon ^2\left(-\frac{1}{24}+\frac{1}{3 n}+\frac{1}{12
   n^2}\right) +\epsilon ^4\left(-\frac{77}{6912}+\frac{17}{432 n}+\frac{17}{64 n^2}\right) +\epsilon ^6\left(-\frac{11063}{3317760}-\frac{157}{46080 n}+\frac{44851}{552960
   n^2}\right) + \ord{\epsilon^{8}}
\end{dmath}

\begin{dmath}
\mathbf{n}=1+\epsilon ^2\left(-\frac{1}{2}-\frac{1}{n}\right) +\epsilon ^4\left(-\frac{1}{72}-\frac{5}{24 n}-\frac{35}{24 n^2}\right) +\epsilon ^6\left(\frac{535}{165888}-\frac{1513}{82944 n}-\frac{20357}{41472
   n^2}\right) + \ord{\epsilon^{8}}
\end{dmath}

The lowest quasinormal mode, with
\begin{eqnarray}
 m(t,z) =m_0(z)+ e^{\Omega t} f(z),\quad p(t,z) = p_0(z) + e^{\Omega t} g(z),
\end{eqnarray}
has frequency (up to NNLO and $\epsilon^{6}$;  results up to 4NLO and $\epsilon^6$ in the \textsl{Mathematica} files)
\begin{eqnarray}
\begin{split}
\Omega =&
   -\epsilon ^2
   \left(\frac{1}{12}-\frac{5}{6 n}-\frac{-18+\pi ^2}{36
   n^2}
   \right)-\epsilon ^4 \left(\frac{19}{576}-\frac{17}{96
   n}-\frac{156+5 \pi ^2}{576 n^2}
   \right)\\
   &-\epsilon ^6 \left(\frac{63317}{4976640}-\frac{54689}{2488320
   n}-\frac{1576074+62597 \pi ^2}{14929920 n^2}
   \right)\,,
   \end{split}\label{Omegafund}
\end{eqnarray}
and
\begin{eqnarray}
\begin{split}
\bm{\Omega} =&
   -\epsilon ^2
   \left(\frac{1}{12}-\frac{5}{6 n}-\frac{-18+\pi ^2}{36
   n^2}
   \right)-\epsilon ^4 \left(\frac{7}{192}-\frac{23}{96
   n}-\frac{36+17 \pi ^2}{1728 n^2}
   \right)\\
   &-\epsilon ^6 \left(\frac{75497}{4976640}-\frac{138869}{2488320
   n}-\frac{683634+73337 \pi ^2}{14929920 n^2}
   \right)\,.
   \end{split}\label{Omegafund}
\end{eqnarray}


\end{document}